\documentclass[12pt]{spieman}  
\usepackage{amsmath,amsfonts,amssymb}
\usepackage{graphicx}
\usepackage{setspace}
\usepackage{tocloft}
\usepackage{comment}
\usepackage{siunitx}
\usepackage{cite}
\usepackage[hang,flushmargin]{footmisc}

\newcommand\blfootnote[1]{%
  \begingroup
  \renewcommand\thefootnote{}\footnote{#1}%
  \addtocounter{footnote}{-1}%
  \endgroup
}

\title{Resampling and super-resolution of hexagonally sampled images using deep learning}

\author[a*]{Dylan Flaute}
\author[a]{Russell C. Hardie}
\author[a]{Hamed Elwarfalli}
\affil[a]{University of Dayton, Department of Electrical and Computer Engineering, Dayton, Ohio, United States}

\cftpagenumbersoff{figure}
\cftpagenumbersoff{table} 

\begin{document} 
\maketitle

\begin{abstract}
Super-resolution (SR) aims to increase the resolution of imagery.  Applications include security, medical imaging, and object recognition.
We propose a deep learning-based SR system that takes a hexagonally sampled low-resolution image as an input and generates a rectangularly sampled SR image as an output.
For training and testing, we use a realistic observation model that includes optical degradation from diffraction and sensor degradation from
detector integration.  Our SR approach first uses non-uniform interpolation to partially upsample the observed hexagonal
imagery and convert it to a rectangular grid.  We then leverage a state-of-the-art convolutional neural network (CNN) architecture designed for SR known as
Residual Channel Attention Network (RCAN).  In particular, we use RCAN to further upsample and restore the imagery to produce the final SR image estimate.
We demonstrate that this system is superior to applying RCAN directly to rectangularly sampled LR imagery with equivalent sample density.
The theoretical advantages of hexagonal sampling are well known.  However, to the best of our knowledge, 
the practical benefit of hexagonal sampling in light of modern processing techniques such as RCAN SR
is heretofore untested. Our SR system demonstrates a notable advantage of hexagonally sampled imagery when employing a modified RCAN for hexagonal SR.
\end{abstract}


{\noindent \footnotesize\textbf{*}Dylan Flaute,  \linkable{dylan.flaute@gmail.com}
\\\\
© 2021 Society of Photo‑Optical Instrumentation Engineers (SPIE). One print or electronic copy may be made for personal use only. Systematic reproduction and distribution, duplication of any material in this publication for a fee or for commercial purposes, and modification of the contents of the publication are prohibited.  
}

\section{Introduction}
\label{sect:intro}

Super-resolution (SR) is a term for the task of increasing the resolution of low-resolution (LR) images.
The SR task has broad applications in security and surveillance \cite{Zou2010VeryLR,Li2018FaceRI}, medical imaging \cite{Shi2013CardiacIS}, and object recognition \cite{Sajjadi2017EnhanceNetSI}.
Single-image SR (SISR) is fundamentally challenging because it generally requires the interpolation of undersampled imagery without access to additional spatial samples from other frames.
When processing LR data acquired directly from a camera, the observation model
includes diffraction blurring and and detector integration effects, in addition to 
potential aliasing that results from undersampling. 
\blfootnote{Cite as: Dylan Flaute, Russell C. Hardie, Hamed Elwarfalli, "Resampling and super-resolution of hexagonally sampled images using deep learning," Opt. Eng. 60(10), 103105 (2021), \url{https://doi.org/10.1117/1.OE.60.10.103105}}


Effectively addressing the ill-posed SISR problem requires the
exploitation of prior statistical information about the images.  
Deep learning is a broad class of data-driven approaches to 
deriving statistical information.
Recently, deep learning-based approaches have pushed the state-of-the-art for inverse problems in image processing, including but not limited to denoising, deconvolution, SR, and inpainting \cite{Ongie2020DeepLT}.
In the field of SISR, the use of neural networks \cite{Davila:00} and convolutional neural networks (CNNs) has been widely successful.
In CNN-based approaches, convolutional filters are learned based on very large datasets of paired LR and high-resolution (HR) images.
Examples of CNN-based approaches to SISR include SRCNN \cite{Dong2016ImageSU}, FSRCNN \cite{Dong2016AcceleratingTS}, SCN \cite{Wang2015DeepNF}, VDSR \cite{Kim2016AccurateIS}, DRRN \cite{Tai2017ImageSV}, LapSRN \cite{Lai2017DeepLP}, MSLapSRN \cite{Lai2019FastAA}, ENet-PAT \cite{Sajjadi2017EnhanceNetSI}, MemNet \cite{Tai2017MemNetAP}, EDSR \cite{Lim2017EnhancedDR}, SRMDNF \cite{Zhang2018LearningAS}, and RCAN \cite{Zhang2018ImageSU}.
A comprehensive survey on deep learning-based SISR approaches is provided in Ref.~\citenum{Wang2020DeepLF}.
Deep learning has also been successful in the field of deblurring.
DeepDeblur \cite{Nah2017DeepMC},  ED-DSRN \cite{Zhang2018ADE}, GFN \cite{Zhang2018GatedFN}, and EDSR \cite{Lim2017EnhancedDR} all approach deblurring, mostly in addition to SR, with deep learning.

Although CNN-based approaches to learning and exploiting prior statistical information have been successful, the approach is still fundamentally bounded by the information present in the samples.
Alternative approaches to advancing the state of SISR examine the sampling.
For example, CAR \cite{Sun2020LearnedID} is a deep learning approach to find a more optimal down-sampling function that is easier for another network to undo.
A simpler, theoretically-grounded sampling alternative is to sample on a hexagonal grid rather than a rectangular grid.
This approach has the theoretical motivation that a hexagonal grid with the same density as the rectangular grid can support a higher cutoff frequency without introducing aliasing.

Some recent works demonstrate the benefits of hexagonal imagery for processing purposes by processing the hexagonal image directly.
For example, the image classification system HexaConv \cite{Hoogeboom2018HexaConv} processes hexagonal images by adapting the Group-equivariant Convolutional Neural Network (G-CNN) architecture \cite{Cohen2016GroupEC} to leverage the additional symmetries that are found on a hexagonal grid, yielding improved classification performance at a given parameter budget than baseline on the natural images in the CIFAR-10 dataset and on the aerial imagery, which is rotation-invariant by nature, in the Aerial Image Dataset (AID) \cite{Xia2017AIDAB}.
The image classification system HexagonNet \cite{Luo2019HexagonalCN} also processes hexagonal images directly and increments performance further on the AID dataset as well as achieving a new state-of-the-art on the ModelNet40 shape classification dataset.
And, the authors of Ref.~\citenum{Zhang2019OrientationAwareSS} use hexagonally-defined filters to process omnidirectional imagery for orientation-aware semantic segmentation.
To facilitate future research into this type of processing, the Python library HexagDLy \cite{hexagdly_paper} extends the deep learning framework PyTorch \cite{NEURIPS2019_9015} to include efficient hexagonal convolutions.
The more abstract problem of developing convolution operators for any non-Cartesian grid is approached by the authors of Ref.~\citenum{Jacquemont2019IndexedOF}, 
who use nearest-neighbor information to index the convolution operations.
Aside from image processing, hexagonally gridded data is also useful for estimating the supply-demand gap for ride-sharing due to the increased symmetries of the grid \cite{Ke2019HexagonBasedCN}.
Joint resampling and restoration of hexagonally sampled images using an 
Adaptive Wiener Filter has also been explored Ref.~\citenum{Burada2015JointRR}.

The theoretical advantages of hexagonal sampling are well known. However, hexagonal sampling comes at a cost. Most display devices and image processing algorithms and software today assume rectangular sample/pixel layouts.  Furthermore, to the best of our knowledge, 
the practical benefit of hexagonal sampling in light of modern SR processing techniques
is heretofore untested.  In this paper, we demonstrate a system for SISR that takes hexagonally sampled LR imagery as inputs and generates rectangularly sampled SR images as outputs.
To make the problem more realistic, our observation model includes the effects of diffraction and detector integration for a given camera system.  Our SR approach first uses non-uniform interpolation to partially upsample the observed hexagonal
imagery and convert it to a rectangular grid.  We then leverage a state-of-the-art CNN architecture designed for SR known as Residual Channel Attention Network (RCAN)\cite{Zhang2018ImageSU}.  In particular, we use RCAN to further upsample and restore the imagery to produce the final SR image estimate.  To enrich the RCAN input, we also supply subpixel interpolation distances for each pixel and feed into the RCAN architecture through a separate input head.
This modification is inspired by the FIFNET architecture for multi-image SR \cite{Elwarfalli2021FIFNETAC}.

Since paired hexagonally sampled LR images and rectangularly sampled HR images are not readily available, we develop an observation model-based approach for generating realistic, full frame RGB, synthetic data. 
Our observation model effectively constitutes 
a ``forward model'' for the inverse problem we wish to solve.
The observation model leverages a camera-specific optical transfer function (OTF) that models diffraction and detector integration (based on either hexagonal or rectangular detectors) for generating our data \cite{Hardie2015ImpactOD,Elwarfalli2021FIFNETAC} as well as a resampling step. 
We apply the observation model to the DIV2K SISR dataset \cite{Agustsson_2017_CVPR_Workshops} to generate data to train our models so as to be ready to process real full frame RGB color camera outputs.  

We demonstrate that our modified RCAN architecture processing data that was originally sampled hexagonally is superior to applying RCAN directly to rectangularly sampled LR imagery with equivalent sample density.
To make this comparison, we choose a sample density such that no interpolation is necessary for generating the rectangularly sampled imagery (and thus, there is no interpolation error).
We generate a rectangularly sampled color-imagery dataset from DIV2K with rectangular detector shape and a hexagonally sampled dataset with hexagonal detector shape.
To restore the rectangular imagery, we train RCAN end-to-end and leverage a pre-trained RCAN model.
Despite having no interpolation error in the input data, the rectangularly-based architecture still underperforms the hexagonally-based architecture.
The implications of our results are two-fold:
First, our system demonstrates the practical benefit of hexagonally sampled imagery for image processing, even in the previously untested context of deep learning-based post-processing.
Second, our system reduces the practical drawback of using hexagonally sampled imagery -- namely, that most display devices and image processing software assumes rectangular sample/pixel layouts -- since our system can be used to resample hexagonal imagery to rectangular grids. 

The remainder of this paper is laid out as follows.
Section~\ref{sec:obs_model} describes the observation model that is used to generate training and testing data.
Section~\ref{sec:approach} describes the architecture of our approach to resampling and restoring the imagery produced by that observation model.
Section~\ref{sec:training_details} describes the experiments we perform to compare hexagonally sampled imagery and rectanguarly sampled imagery post-processed with those architectures.
Section~\ref{sec:exper_results} describes the results of those experiments.
Finally, Section~\ref{sec:ccl} provides our discussion of those results.

\section{Observation Model}\label{sec:obs_model}

\begin{figure}
    \centering
    \includegraphics[clip=true, trim = 1in 2in 1in 2in, width=5.5in]{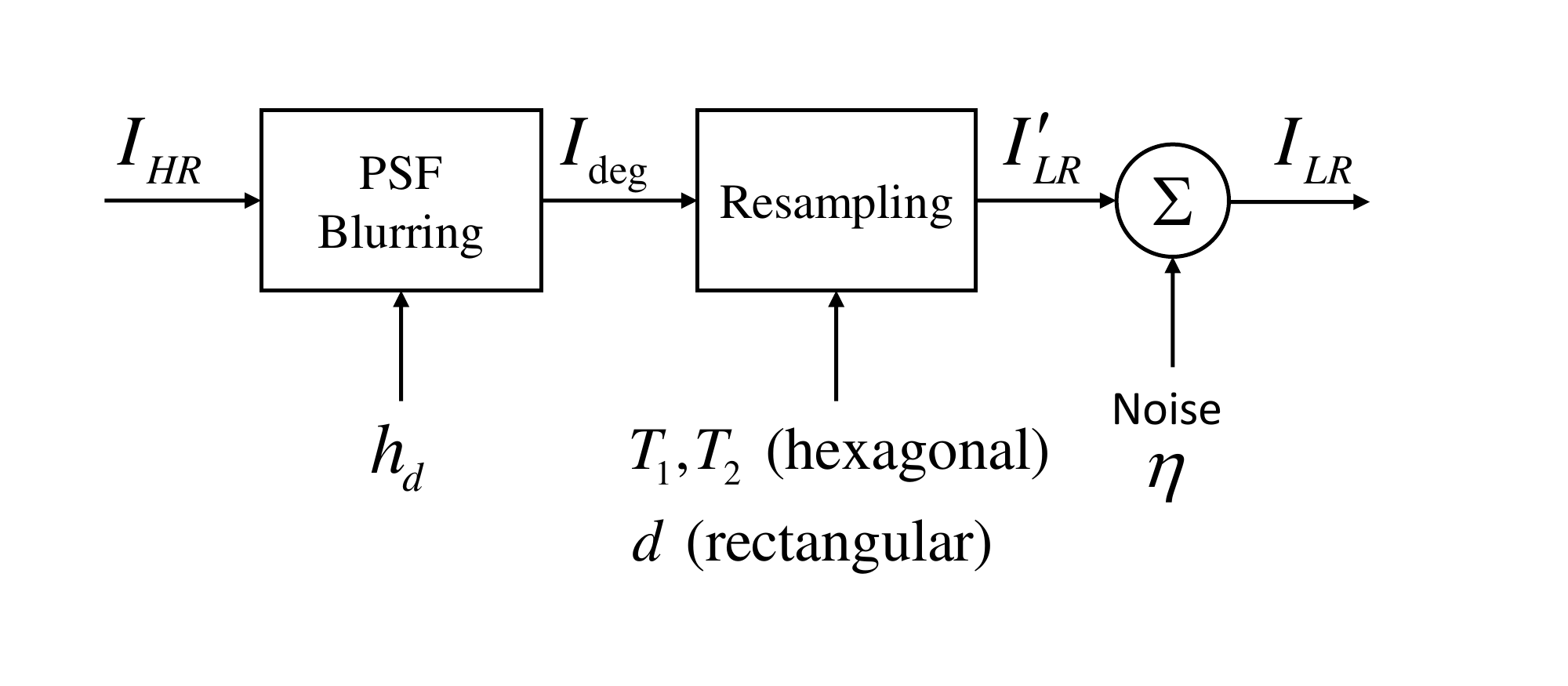}
    \caption{Discrete forward observation model used to provide realistic degradation of ideal images for training and testing purposes.  A pristine HR image is blurred with an impulse-invariant PSF and then resampled according the detector sampling pattern.  Finally, noise is added to produce the LR image.}
    \label{block_diagram}
\end{figure}

A block diagram of our 
observation model is shown in Fig.~\ref{block_diagram}.
The model takes an ideal discrete-space HR image as input 
and outputs a simulated LR image that approximates a camera sensor output.
The observation model consists of two main parts: an optical transfer function (OTF) to account for camera optics and detector integration,
and a resampling step to simulate the LR sampling pattern of the sensor.   We simulate both 
hexagonal and rectangular sampling by resampling the blurred HR image onto the appropriate grid.
Finally, our model includes independent and additive white Gaussian noise with standard deviation $\sigma_\eta$. 
The key parameters of our observation model are listed in Table~\ref{tab:optics} for the experimental study presented here.
The parameters in Table~\ref{tab:optics} have been selected to represent a typical acquisition scenario with an RGB camera, with a few liberties taken for numerical convenience.  The wavelengths correspond to the Hamamatsu RGB Color Sensor S9702.  The f-number is moderate and commonly used for camera lenses.  The simulated rectangular detector pitch is within the range of currently available commercial sensors.  We selected the HR pitch so that the rectangular detector pitch would be an integer multiple, to allow for simple downsampling in modeling the rectangular detector scenario.   The hexagonal pitches were selected to give the same sampling density as the rectangular sampling scenario.  Finally the noise standard deviation has been selected to give a high, but realistic, signal-to-noise ratio (SNR) of approximately 36 dB for the simulated images.  A high SNR was selected because the proposed study is intended to focus on detector sampling and not noise reduction.
Note that we perform the simulation on three color channels for RGB images to model a full frame RGB color camera and some of the parameters in Table~\ref{tab:optics} are wavelength dependent. 
In the following two subsections we shall provide further details on the OTF and resampling blocks, respectively.

\begin{table}
    \caption{Summary of observation model parameters for the red, green, and blue color channels.}
    \label{tab:optics}
    \centering
    \begin{tabular}{|l|c|c|c|}
    \hline
        Parameter & \multicolumn{3}{c|}{Wavelength}  \\
         \cline{2-4}
           & $\lambda_R = 620.0 ~\si{\nano \meter}$ & $\lambda_G = 540.0 ~\si{\nano \meter}$  & $\lambda_B = 460.0 ~\si{\nano \meter}$  \\
                  \hline
        F-number             & $F=4.000 $ & $F=4.000 $ &  $F=4.000 $  \\ 
        Cutoff frequency & $\rho_c=0.4032$ cyc/$\mu$m & $\rho_c=0.4630$ cyc/$\mu$m &  $\rho_c= 0.5435$ cyc/$\mu$m \\ 
        Nyquist pitch & $p_N = 1.240$ $\mu$m & $p_N =1.080 $ $\mu$m &  $p_N = 0.9200$ $\mu$m \\
        HR pitch & $p = 1.000$  $\mu$m & $p = 1.000$ $\mu$m & $p = 1.000$  $\mu$m \\
        Rectangular LR pitch & $d = 4.000$ $\mu$m & $d = 4.000$ $\mu$m & $d = 4.000$ $\mu$m  \\     
        Hexagonal LR pitch 1 & $T_1 = 4.298$ $\mu$m & $T_1 = 4.298$ $\mu$m & $T_1 = 4.298$ $\mu$m  \\   
        Hexagonal LR pitch 2 & $T_2 = 7.445$ $\mu$m & $T_2 = 7.445$ $\mu$m & $T_2 = 7.445$ $\mu$m  \\   
        Noise std. & $\sigma = 1.000$ DU & $\sigma = 1.000$ DU & $\sigma = 1.000$ DU \\
        \hline
        \hline
    \end{tabular}
\end{table}

\subsection{Optical transfer function}\label{otfsec}

In this section we discuss the OTF model that accounts for diffraction-limited optics 
and blurring from spatial integration of the detector elements.
Together, these two components can be combined to provide the overall OTF model as
\begin{equation}\label{eq:otf}
    H(u,v) = H_{dif}(u,v) H_{det}(u,v),
\end{equation}
where $u$ and $v$ are the horizontal and vertical spatial frequencies in cycles per unit distance.
The first term, $H_{dif}$, is the OTF for diffraction-limited optics and the second term, $H_{det}$, accounts for detector integration.

The OTF component for diffraction assuming a circular pupil function \cite{goodman2005introduction}
is given by
\begin{equation}
    H_{dif}(\rho) = \begin{cases} 
        \frac{2}{\pi} \left[ \cos^{-1}(\rho)-\rho \sqrt{1-\rho^2} \right] & \rho < 1, \\
        0 & else
    \end{cases}
    \label{eq:Hdif}
\end{equation}
where $\rho = \sqrt{u^2 + v^2}/\rho_c$, $\rho_c = 1/ (\lambda F)$ is the optical cutoff frequency, $\lambda$ is the wavelength of light, and $F$ is the f-number of the optics.  Note that the f-number is defined as the ratio of the focal length of the optics to the effective aperture \cite{Boreman1998BasicEF}.

\begin{figure}[tb]
    \centering
    \includegraphics[clip=true, trim = 1in 3.5in 1in 3.5in, width=5.5in]{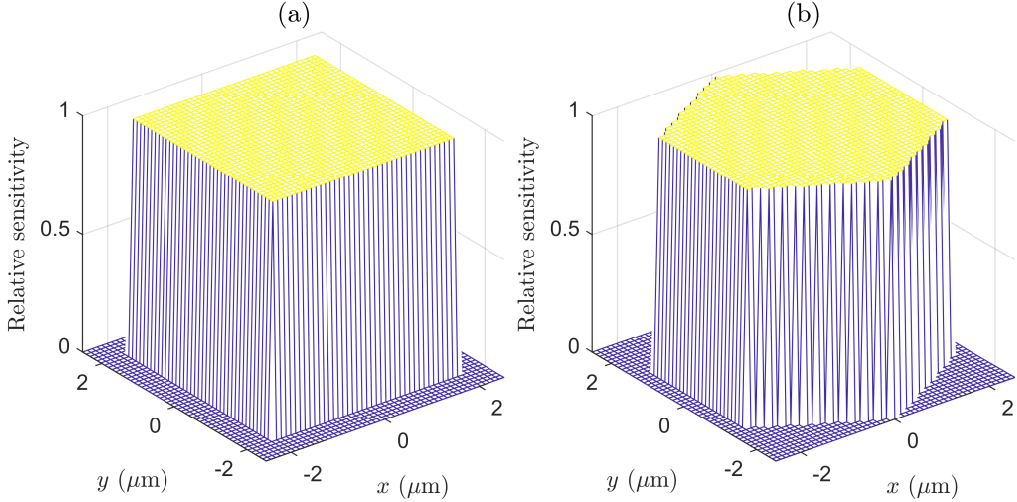}
    \caption{Detector shapes based on the parameters in Table~\ref{tab:optics} for (a) rectangular detectors and (b) hexagonal detectors.}
    \label{fig:detector_shapes}
\end{figure}

The detector component of the OTF accounts for the finite size of the detectors in the camera's focal plane array.
At the native pixel scale, the detector integration plays only a small role in degrading the observed imagery.  However, 
when upsampling an image for SR, the native detector active area will now span multiple HR pixels and play an important and 
potentially dominant role in the observation model.  
The detector OTF can be found by simply taking the 2D Fourier transform of the detector element shape as described 
by Hardie et al\cite{Hardie2015ImpactOD}.  In our study, we wish to present a fair comparison between rectangular and hexagonal sampling.
To do so, we have elected to make the active area of both types of detectors
the same and have 100\% fill factor. This will give rise to the same signal-to-noise ratio
using comparable image acquisition parameters.  
Furthermore, this arrangement gives rise to the same sampling density in spatial 
samples per unit area. 
The spatial detector active areas used for the rectangular and hexagonal detectors are 
shown in Fig.~\ref{fig:detector_shapes}.  The specific dimensions of the detectors shown are based on the parameters in Table~\ref{tab:optics}.

\begin{figure}[p]
    \centering
    \includegraphics[clip=true, trim = 1in 3.5in 1in 3.5in, width=4.25in]{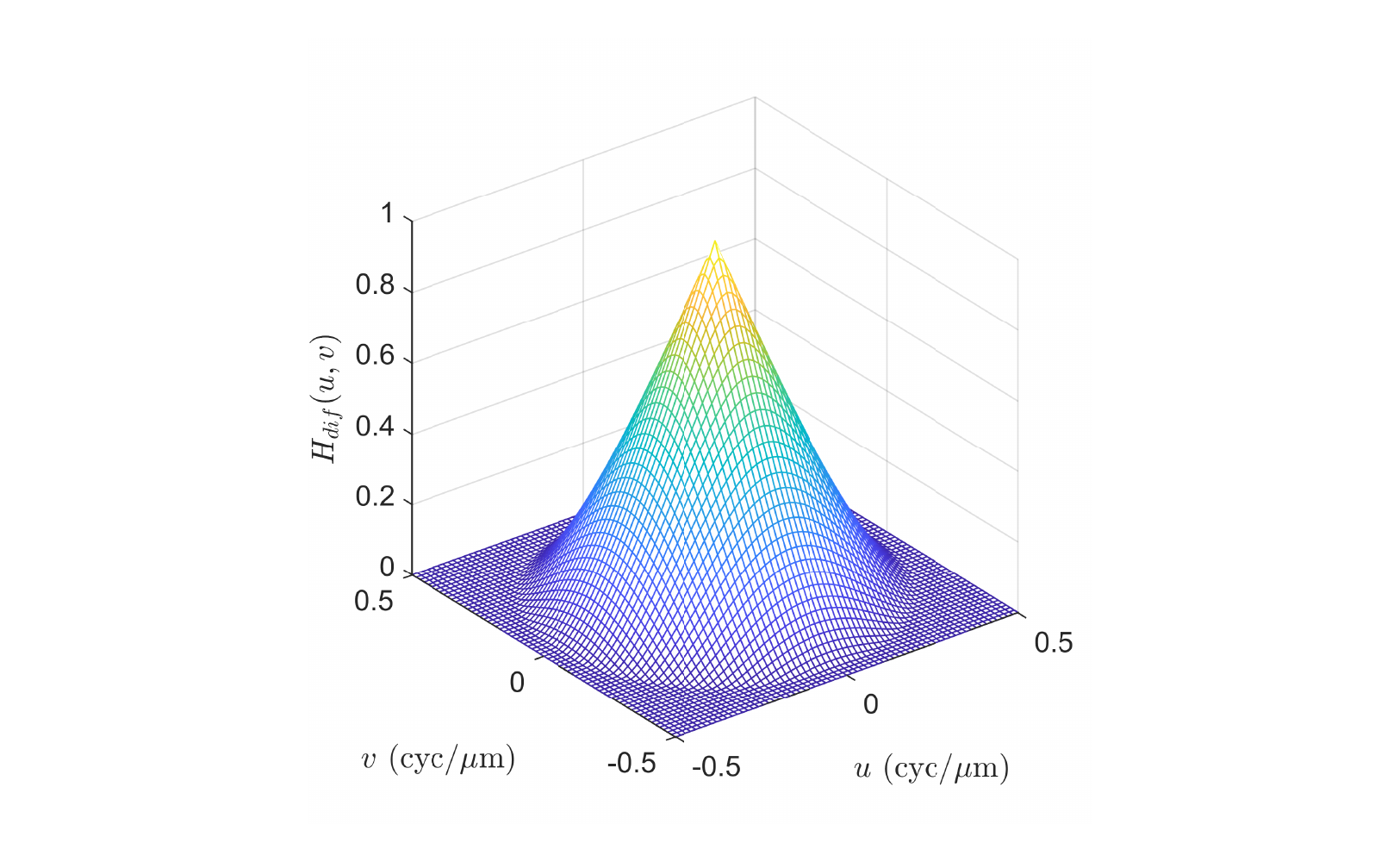}
    \caption{Diffraction OTF model from Eq.~\eqref{eq:Hdif} for the green channel parameters in Table~\ref{tab:optics}.}
    \label{fig:diffraction_otf}
    \centering
    \includegraphics[clip=true, trim = .5in 1.5in .5in 1.5in, width=4.75in]{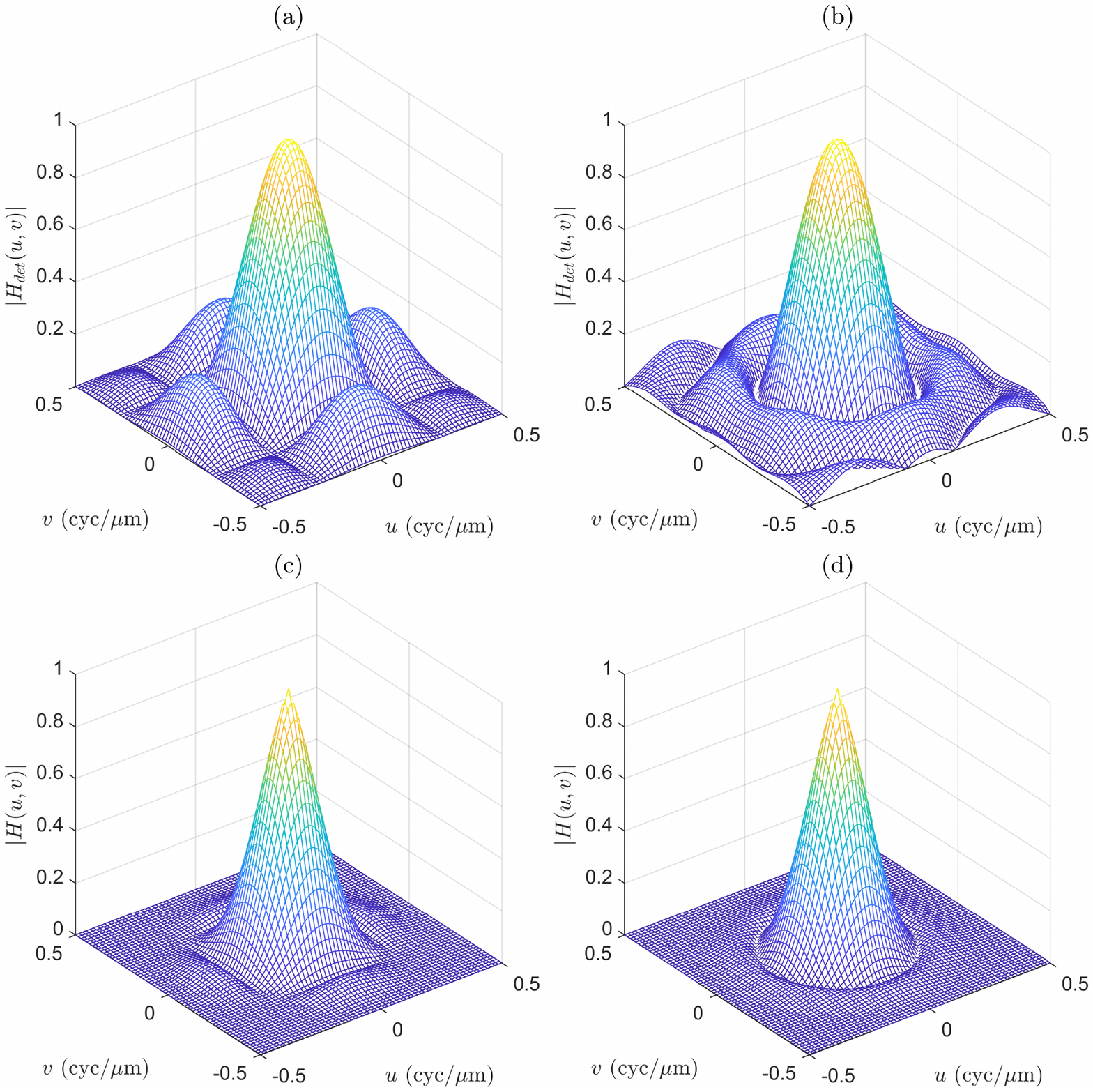}
    \caption{Detector OTFs from for (a) rectangular detectors and (b) hexagonal detectors.   Combined OTFs from Eq.~\eqref{eq:otf} for (c) rectangular detectors and (c) hexagonal detectors. }
    \label{fig:detector_otf4}
\end{figure}

The diffraction OTF is shown in Fig.\ \ref{fig:diffraction_otf} for the parameters in Table~\ref{tab:optics} for the green channel.
The detector OTFs along with the combined OTFs are shown in
Fig.~\ref{fig:detector_otf4} for the green channel.  Note that the detector OTF in Fig.~\ref{fig:detector_otf4}(a) has a sinc form
corresponding to the Fourier transform of the square detector shape in Fig.~\ref{fig:detector_shapes}(a).
and note that the OTF for the hexagonal detector in Fig.~\ref{fig:detector_otf4}(b) has an 
approximately circularly symmetric form by virtue of the hexagonal detector shape shown in Fig.~\ref{fig:detector_shapes}(b).
The combined OTFs that include the diffraction and the detector are shown in Figs.~\ref{fig:detector_otf4}(c) and (d) for rectangular and hexagonal, respectively. 
One can see from the OTFs that the low-pass effect of the 
two different detector types of the same area is quite similar.  
However, one might argue that the more radially symmetric OTF shown in Fig.~\ref{fig:detector_otf4}(d) is somewhat advantageous: it gives an imager that employs hexagonal sampling a more angle-independent resolution performance.

The OTF model in Eq.~\eqref{eq:otf} is for a continuous-space image.  The continuous-space PSF is given by the inverse continuous-space Fourier transform (ICSFT) of the OTF in Eq.~\eqref{eq:otf}.  
This may be expressed as
\begin{equation}
    h(x,y) = \text{ICSFT}\{ H(u,v) \}.
\end{equation}
To process the discrete high-resolution input frames, as shown in our observation model in Fig.~\ref{block_diagram}, we employ the
impulse-invariant discrete-space equivalent blurring.  To do so, we assume our pristine input images are sampled at the Nyquist rate with a pitch of $p$. The Nyquist criterion dictates that $1/p > 2 \rho_c = 2/(\lambda F)$.  Equivalently, this means we must have $p < \lambda F / 2$ to guarantee no aliasing \cite{Hardie2015ImpactOD}.
An impulse-invariant equivalent discrete-space PSF, $h_d(n_1,n_2)$, may be found by sampling $h(x,y)$ with a sample spacing of $p$.  This impulse-invariant PSF is applied to produce
\begin{equation}
    I_{deg}(n_1,n_2) = I_{HR}(n_1,n_2) * h_d(n_1,n_2),
\end{equation}
where $*$ is discrete 2D convolution.

To accommodate full color RGB images, we compute the impulse-invariant PSF for the center wavelengths 
of red, green, and blur color channels.  The assumed wavelengths are listed in Table~\ref{tab:optics}.  Note that for our simulation parameters the HR grid rate exceeds the
Nyquist rate for the red and green channels, but is is slightly below Nyquist for the blue channel.
To address this, we truncate the theoretical blue OTF model to prevent any aliasing with regard to the impulse invariant filter. 
To justify this, note that there is an exceedingly small amount of signal energy beyond the folding frequency that is truncated. In particular, we computed the total volume of the OTF and the volume inside the folding frequency of the 1 $\mu$m HR sample spacing for the blue channel. The OTF volume beyond the folding frequency (0.5 cyc/$\mu$m) represents only 0.0105\% of the total OTF volume (or a fraction of 0.000105). Visually, note that the blue OTF is very similar to the green OTF shown in Fig. 4(c) and (d), and the OTF values in the figure are very small near the $0.5$ cyc/$\mu$m folding frequency. Based on this,
we believe this small adjustment to the blue OTF model is 
negligible and will not adversely impact the fidelity of the study.

\subsection{Spatial sampling}\label{samplesec}

\begin{figure}[tb]
    \centering
    \includegraphics[clip=true, trim = 2in 3in 2in 3in, width=3.5in]{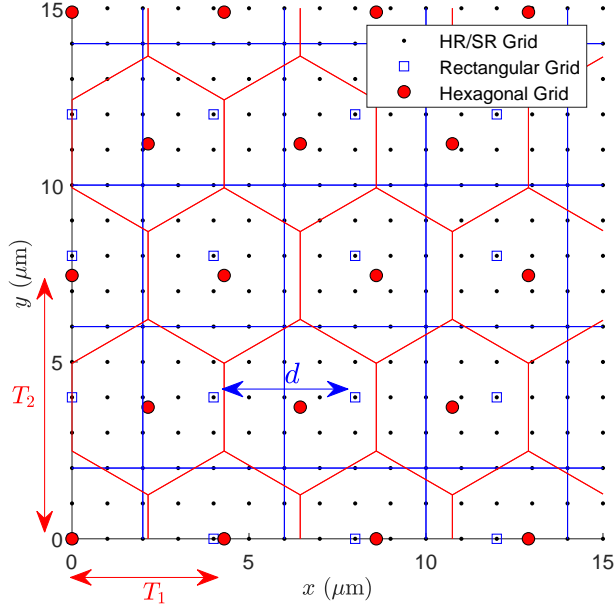}
    \caption{Sampling grids used in the observation model.  The HR grid (black) is a dense rectangular Nyquist sampling configuration.  The LR rectangular sampling grid and associated detectors are shown in blue.  The LR hexagonal sampling grid and associated detectors are shown in red.}  
    \label{fig:hex_grid2}
\end{figure}

In the previous section, we outlined the PSF blurring model that simulates camera optics and detector integration on the HR image.  The next step of the observation model is spatial resampling of the blurred input image $I_{deg}$ according to the corresponding detector layout.  The sampling grids using the parameters listed in Table~\ref{tab:optics} are shown in Fig.~\ref{fig:hex_grid2}.  The HR grid (black) is a dense rectangular Nyquist sampling configuration with spacing $p=1$ $\mu$m.  The pristine input image, $I_{HR}$, is assumed to be sampled on this grid, as is the SR output image.  The LR rectangular sampling grid (blue) has a sample spacing of $d$.  In our current experiment we have $d=4p$.  The hexagonal LR grid (red) can be thought of as two interlaced rectangular grids with spacing of $T_1$ in the horizontal dimension and $T_2$ in the vertical dimension.  The interlacing offset is effectively a 1/2 sample spacing shift in the horizontal and vertical dimensions.  To form an ideal hexagonal sampling pattern, we require that $T_2=\sqrt{3}T_1$ \cite{Middleton2005HexagonalIP}.  Approximate hexagonal grids are also 
sometimes employed \cite{Burada2015JointRR} where the relationship between $T_2$ and $T_1$ is a rational fraction, or even simply a 2-to-1 grid where $T_2=2 T_1$.  These approximately hexagonal grids have less perfect frequency packing properties, but may provide other benefits with regard to fabrication or post processing.

Based on the LR sample positions, we use bicubic interpolation
on the image $I_{deg}$ to produce $I^{\prime}_{LR}$.  Since $I_{deg}$ is on a Nyquist grid, the downsampling interpolation here is very accurate and unhindered by aliasing. We then add independent white Gaussian noise with a standard deviation of $\sigma_\eta=1$ digital unit (DU).  Finally, we quantize the intensity to $2^8$ levels to form the final LR image, denoted $I_{LR}$.  This process is repeated for each of the color channels.  Note that we are simulating a full frame RGB color camera, and not color filter mosaic based images. 
The spatial sampling density of the LR rectangular sampling grid is $1/d^2$.   For the LR hexagonal grid the sampling density is $2/(T_1T_2)=2/(\sqrt{3}T_1^2)$.  By setting the sampling densities of the rectangular and hexagonal patterns equal, we obtain the relationship
\begin{equation}
T_1 = \sqrt{  \frac{2}{\sqrt{3}}  } d.
\label{eq:T1d}
\end{equation}
The relationship in Eq.~\eqref{eq:T1d} is used to set the LR hexagonal grid in relation to the LR rectangular grid in Table~\ref{tab:optics}.

\begin{figure}[tb]
    \centering
    \includegraphics[clip=true, trim = .5in 3.75in .5in 3.75in, width=6.5in]{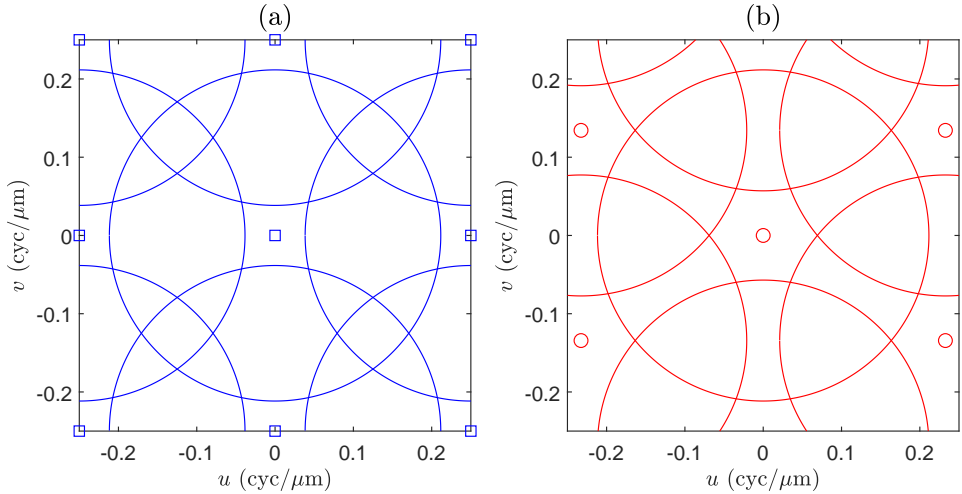}
    \caption{Frequency domain spectral packing using the parameters in Table~\ref{tab:optics} for (a) rectangular sampling and (b) hexagonal sampling.  The circles represent isocontours of the OTF at 10\% of the peak value. }
    \label{fig:freq_pack}
\end{figure}

As we saw in Sec.~\ref{otfsec}, the detector OTF for a hexagonal detector is not that much different from
a comparably sized rectangular detector.  The real benefit of hexagonal sampling is the result of efficient frequency packing and reduced aliasing.  Isotropic rectangular sampling causes the continuous frequency spectrum to repeat in a rectangular pattern every $1/d$ in the horizontal and vertical frequency axis.  With hexagonal sampling, 
the continuous frequency spectrum repeats in a hexagonal pattern.  The repetition occurs every $2/T_1$ along the horizontal frequency axis and every $2/T_2$ along the vertical frequency axis. 
That pattern is itself repeated with an offset of $1/T_1$ and  $1/T_2$ along the respective frequency axes.  The frequency packing for the parameters in Table~\ref{tab:optics} is shown in
Fig.~\ref{fig:freq_pack}.  The circles represent the isocontours of the OTF at a level of 10\% of the maximum.   Clearly the overlapping circles show that there is a large amount of potential aliasing in both the rectangular and hexagonal sampling cases.  However, the intrusion of the repeating spectra around DC frequency at $u=v=0$ cycles/$\mu$m is demonstrably less in the case of the hexagonal sampling in 
Fig.~\ref{fig:freq_pack}(b) compared with rectangular in Fig.~\ref{fig:freq_pack}(a).

We can quantify the improved hexagonal frequency packing two different ways.
First, consider that to sample exactly at the Nyquist rate with an isotropic rectangular sampling grid, we require $d = 1/(2 \rho_c)$.  For a hexagonal sensor, the equivalent requirement can be shown to be $T_1 = 1/(\sqrt{3} \rho_c)$, based on the frequency domain geometry.  We can combine these relationships with those for sampling density.  This tells us that the sampling density of a rectangular grid for Nyquist sampling is
\begin{equation}
S_{rect} = \frac{1}{d^2} = 4 \rho_c^2.
\label{eq:Srect}
\end{equation}
The equivalent Nyquist sampling density for hexgonal sampling is given by
\begin{equation}
S_{hex} = \frac{2}{ \sqrt{3}T_1^2} = \frac{6\rho_c^2}{ \sqrt{3}  }.
\label{eq:Shex}
\end{equation}
Using these equations, we see that for the same $\rho_c$, the hexagonal spatial sampling grid can be $13.40\%$ less dense.
Another way to describe the potential benefits of hexagonal sampling is to
consider the different cut-off frequency each sampling method can accommodate with no aliasing given the same sampling density.  Performing this calculation using Eqs.\ \eqref{eq:Srect} and \eqref{eq:Shex}, it can be shown that hexagonal sampling can accommodate a cut-off frequency that is $7.46\%$ larger than rectangular.
These theoretical advantages of hexagonal sampling are well known.  However, to the best of our knowledge, the practical benefit of hexagonal sampling in light of modern processing techniques such as RCAN SR is heretofore untested. Thus, we believe it is very interesting to see if
CNN based SR processing can exploit the advantages of hexagonal sampling in a significant way.

\section{Super-Resolution for Hexgonally Sampled Images}\label{sec:approach}

In this section, we describe our approach for resampling and restoring rectangular and hexagonal imagery.  Both approaches rely on the RCAN architecture \cite{Zhang2018ImageSU} for SR.
We briefly describe the RCAN architecture before defining the rectangular and hexagonal system.

\subsection{RCAN architecture}\label{sec:rcan_arch}

\begin{figure}
    \centering
    \includegraphics[width=\textwidth]{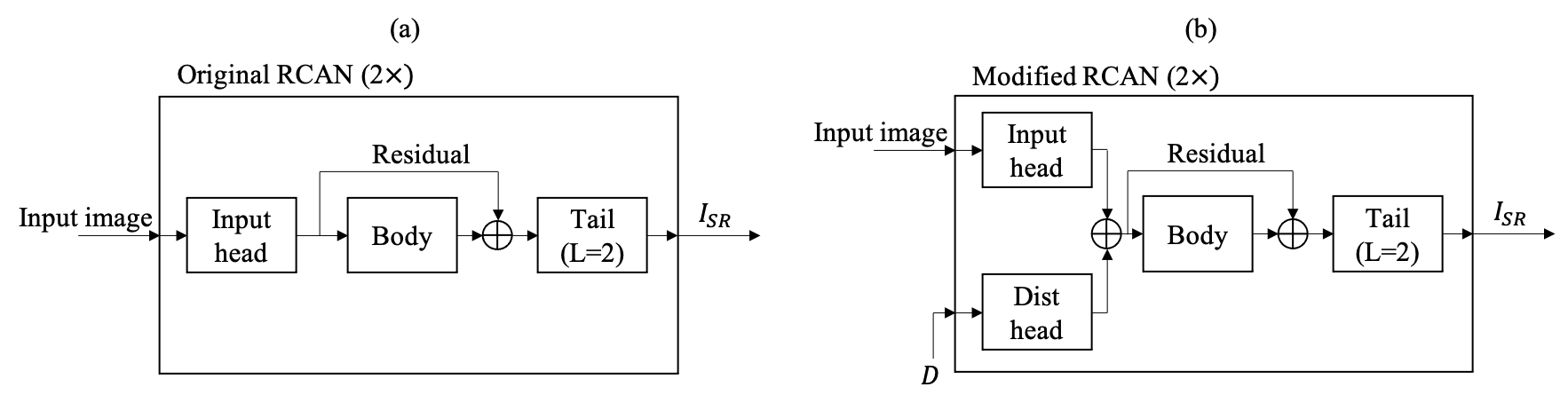}
    \caption{The RCAN architecture for $2 \times$ upsampling. (a) The original architecture. (b) Modified architecture that adds a subpixel distance array to aid in hexagonal image SR.}
    \label{fig:rcan_arch}
\end{figure}

The RCAN architecture is a convolutional SR architecture for integer scale factors.
It consists of stacks of learned convolutional filtering layers and upscaling.
The architecture varies slightly depending on scale factor.
When relevant, the architecture that performs $L\times$ upsampling will be notated as RCAN ($L \times$).
A high-level diagram of RCAN ($2 \times $) is shown in Fig.~\ref{fig:rcan_arch}(a).
Our modified RCAN architecture is shown in Fig.~\ref{fig:rcan_arch}(b), which we shall explain later in Sec.~\ref{sec:sub_pixel_distance}. 

We denote the input of RCAN generally as $I_{LR}$.
This input consists of the LR three-channel RGB color images. 
The standard input head of RCAN consists of a single convolutional layer that extracts 
feature channels from the input.  That is,
\begin{equation}\label{eq:f0}
    F_0 = H_{\text{input}}(I_{LR}),
\end{equation}
where $H_{\text{input}}$ denotes the learned convolutional layer and $F_0$ denotes the learned ``shallow feature.''
The body of RCAN uses Residual-In-Residual (RIR) deep feature extraction.
The RIR body consists of $G$ ``residual groups'' followed by a ``long skip connection'' (LSC).
Denoting the $g$th residual group as $H_g$ ($1 \leq g \leq G$), we can formulate the RIR body as
\begin{equation}
    F_{DF} = F_0 + \text{Conv}_{LSC} \left[ H_G(H_{G-1}( \cdots H_1 (F_0) \cdots )) \right],
\end{equation}
where $\text{Conv}_{LSC}$ is a convolutional layer at the tail of the RIR.

Each residual group $H_g$ is composed of $N_{(g)}$ Residual Channel Attention Blocks (RCABs), a single convolutional layer, and a residual connection.
Each RCAB first performs learned convolutions and then performs channel attention.
Channel attention adaptively rescales channels in the feature map to exploit interdependencies across channels and emphasize important feature channels.
The output of the channel attention mechanism is residually connected to the input of the RCAB.
It is claimed that the channel attention mechanism allows the network to emphasize more useful features and makes the network more discriminative \cite{Zhang2018ImageSU}.

The tail of the RCAN architecture consists of an upsampling module and a final restoration module.  Note that the output is a three-channel rectangularly sampled RGB SR image. 
For upsampling, the RCAN architecture avoids the transposed convolution layers seen in encoder-decoder architectures \cite{Dong2016AcceleratingTS} and opts for the light-weight ESPCN/PixelShuffle strategy \cite{Shi2016RealTimeSI}.
An additional convolutional layer is used on the output.
So, the SR image is formed as
\begin{equation}\label{eq:i_sr}
    I_{SR} = \text{Conv}_{out} (H_{up} (F_{DF})),
\end{equation}
where $H_{up}$ denotes the upsampling module and $\text{Conv}_{out}$ denotes the output convolutional layer.

\subsection{Hexagonal inverse model}\label{sec:hex_inv_model}

\begin{figure}
    \centering
    \includegraphics[width=\textwidth]{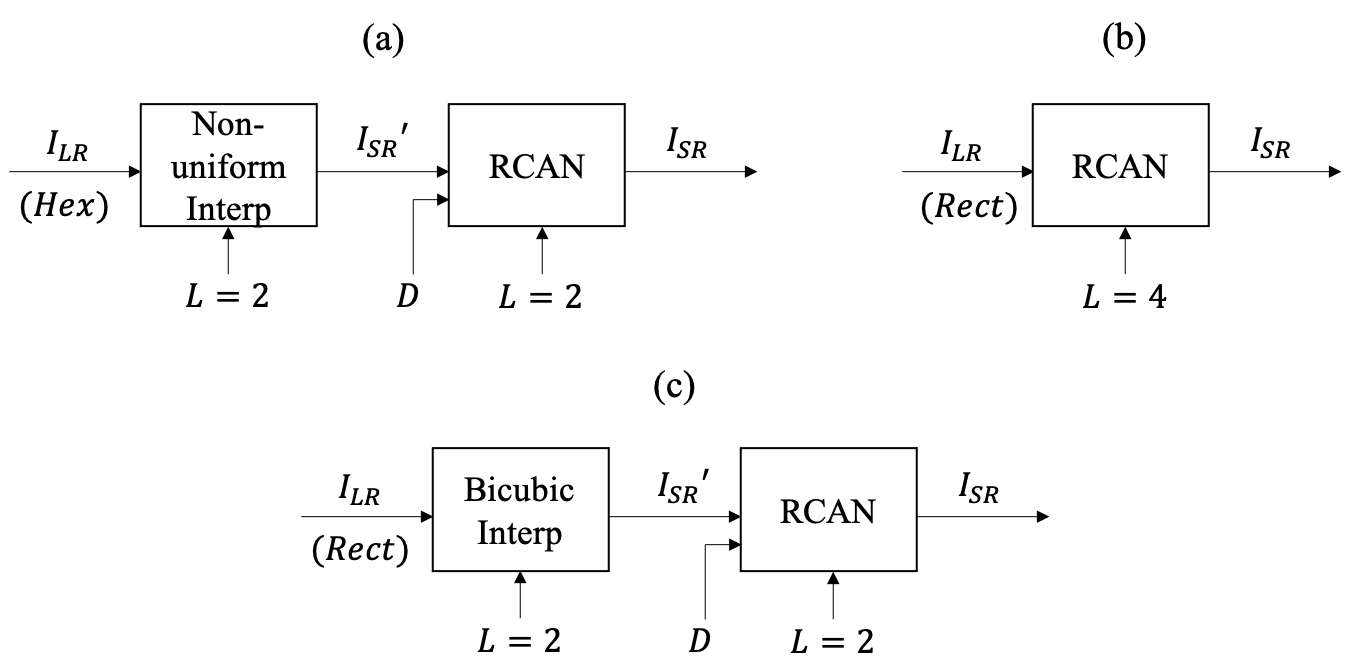}
    \caption{(a) The two-stage hexagonal SR system with LR hexagonal input and SR output. 
    (b) Standard RCAN ($4 \times$) processing used for LR rectangular imagery. (c) The two-stage rectangular SR system.} 
    \label{fig:flowchart}
\end{figure}

For processing hexagonally sampled imagery, we develop a system that uses 
nonuniform interpolation to partially upsample the observed hexagonal
imagery and convert it to a rectangular grid, and then feeds this imagery into an RCAN ($2 \times$)-based network.
The overall proposed system is shown in Fig.~\ref{fig:flowchart}(a). 
As references, we also apply the standard RCAN ($4 \times$) architecture to rectangularly sampled imagery 
with the same sampling density as shown in Fig.~\ref{fig:flowchart}(b) and we apply a directly comparable rectangular version of our system to the same rectangularly sampled imagery as shown in Fig.~\ref{fig:flowchart}(c). 

To motivate our system, ideally we would feed the LR hexagonal data directly into RCAN, and then RCAN would
jointly resample and restore the image. Unfortunately, this is not practical with hexagonal imagery since the upsampling module in RCAN is designed to input and output images with the same aspect ratio and images on a hexagonal grid do not have the same ``aspect ratio'' as images on a rectangular grid (in terms of pixel counts).
It is non-trivial to modify the upsampling module in RCAN to handle different aspect ratios and we are not aware of any existing methods to this end.
Thus, we cannot simply feed the hexagonal image into the network and learn the inverse function.
A naive solution would be to resample the hexagonal imagery onto a rectangular grid of the same sample density as the input. However, this solution would lose the extra spatial frequency content in the hexagonal samples.

Our system addresses both of these problems by using a two-step approach.
In the first step, we use nonuniform interpolation to upsample $I_{LR}$ onto a rectangular grid with twice the input image's sample density.
We call the upsampled image $I_{SR}'$.
The sample spacing of $I_{SR}'$ is $d/2$.
Increasing the sampling density in this step increases the cutoff frequency of the grid and is sufficient to encode the frequency content of the hexagonal image (as limited by the hexagonal grid structure's cutoff frequency).
In the second step, $I_{SR}'$ is fed into the trained RCAN ($2 \times$) network described in Sec.~\ref{sec:rcan_arch} and illustrated in Fig.~\ref{fig:rcan_arch}(b).
We train the network on $(I_{SR}',I_{SR})$ pairs to upsample $I_{SR}'$ onto a rectangular grid with twice the sample density of $I_{SR}'$ (i.e., a sample spacing of $d/4$) while simultaneously correcting for interpolation error introduced in the first step of our system.
The training process is described in detail later, in Sec.~\ref{sec:training_details}.
The output of the system overall is $I_{SR}$, which is the output of the RCAN as defined by Eq.~\eqref{eq:i_sr}.

\subsubsection{Sub-pixel distance matrix}\label{sec:sub_pixel_distance}

The nonuniform interpolation step used by the method in Fig.~\ref{fig:flowchart}(a)
creates interpolation error that varies spatially.  Some interpolated pixel locations are
located very near the original hexagonal data and others are much farther.  
Inspired by the FIFNET sub-pixel registration channel \cite{Elwarfalli2021FIFNETAC}, we incorporate a sub-pixel distance matrix input to the network to quantify the alignment between the hexagonally sampled output of the observation model and the interpolated image that is fed into our modified RCAN shown in Figs.~\ref{fig:flowchart}(a) and \ref{fig:rcan_arch}(b).
This aids our modified RCAN system in dealing with the 
spatially varying interpolation error.  

\begin{figure}
    \centering
    \includegraphics[width=0.6\textwidth]{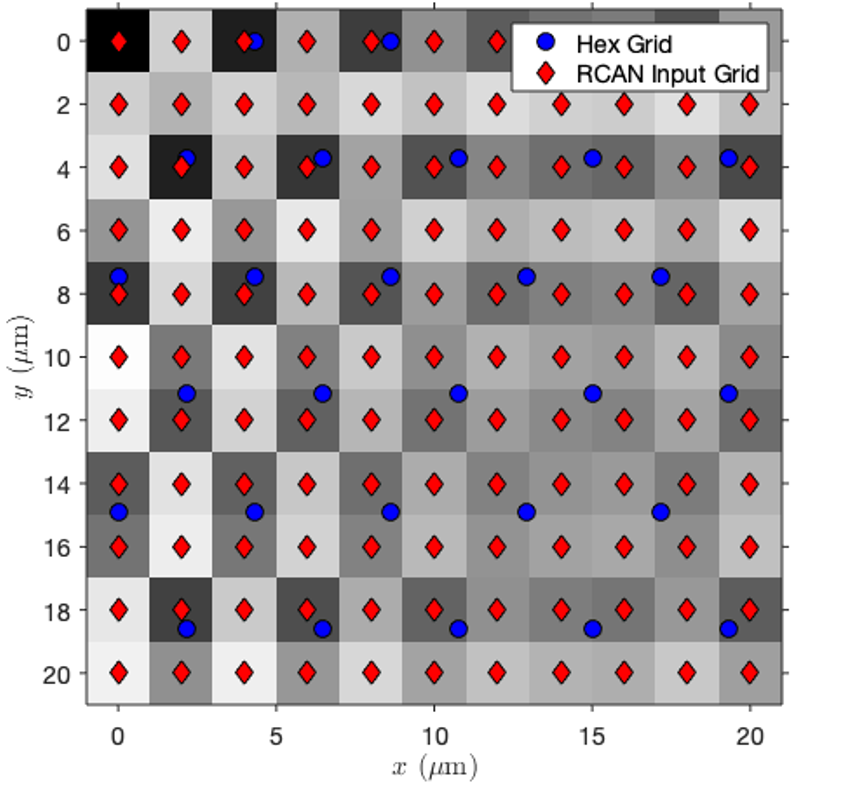}
    \caption{Sample subpixel distance matrix with the hexagonal grid and interpolation grid overlaid.  The hexagonal sampling grid, $X$, is shown in blue.  The rectangular sampling grid is shown in red.  The distance array, $D$, is represented by the pixel grayscale values.}
    \label{fig:subpixel_dist_matrix}
\end{figure}

A formalization of the distance matrix follows.
Let $X$ be the set of spatial coordinates on the hexagonal lattice with elements
${\bf x}_{i.j} \in \mathbb{R}^2$.  Similarly, let the spatial coordinates of the samples in
the $M \times N$ non-uniformly interpolated image be denoted ${\bf r}_{m,n} \in \mathbb{R}^2$.
Then the distance matrix $D \in \mathbb{R}^{M \times N}$ is defined such that
\begin{equation}
D_{m,n} = \mathop {\min }\limits_{{\bf x}_{i,j}\in X} 
{\left\| {\bf r}_{m,n} - {\bf x}_{i,j} \right\|_2}.
\end{equation}
In other words, $D_{m,n}$ encodes how close the pixel $m,n$ in the non-uniformly 
interpolated image is to the nearest sample in the hexagonally sampled input.
A sample of the subpixel distance matrix is shown in Fig.~\ref{fig:subpixel_dist_matrix}.  
The hexagonal
sampling grid, $X$, is shown in blue.  The interpolated sampling grid is shown
in red.  The distance value for every interpolated sample, $D_{m,n}$, is indicated by the
grayscale of the pixel behind the red sample.  Note that red samples that are located
near a blue sample have a small distance value (i.e., dark pixel).  We expect
these red samples to have less interpolation error.  The opposite is true
for red samples located far form the nearest blue sample.

The distance matrix $D$ is fed into RCAN via a separate input head.
This separate input head consists of a stack of convolutional layers with a large receptive field.  
The output of these convolutional layers is connected to the output of the traditional input head such that Eq.~\eqref{eq:f0} becomes
\begin{equation}
    F_0 = H_{\text{input}}(I_{LR}) + H_{\text{dist}}(D),
\end{equation}
where $H_{\text{dist}}$ denotes the head for the sub-pixel distance matrix input.
The precise architecture of $H_{\text{dist}}$ is described in Table~\ref{tab:dist_head}.
The convolutions are padded and biases is used.
Note that $D$ is shown as an input to the system in the system flowchart in 
Fig.~\ref{fig:flowchart}(a), and the distance head is shown in our modified RCAN architecture illustrated in Fig.~\ref{fig:rcan_arch}(b).

\begin{table}[ht]
    \caption{Distance matrix head architecture.}
    \label{tab:dist_head}
    \centering
    \begin{tabular}{c|c|c}
        Channels In & Channels Out & Filter Size \\
        \hline
        $1$ & $64$ & $7 \times 7$ \\
        $64$ & $64$ & $7 \times 7$ \\
        $64$ & $64$ & $7 \times 7$ \\
    \end{tabular}
\end{table}

\section{Experimental Setup}\label{sec:training_details}

In this section, we describe our methodology for comparing hexagonally sampled imagery and rectangularly sampled imagery with deep learning-based post-processing for super-resolution.
We are interested in comparing our proposed system for processing hexagonally sampled imagery with systems for processing rectangularly sampled imagery for super-resolution in order to demonstrate the benefits of hexagonal sampling.
We also consider several baseline systems.
\begin{itemize}
    \item \textbf{Hex + NI ($\mathbf{4\times}$).} For hexagonally sampled data, we use nonuniform interpolation (NI) to upscale by a factor of 4 (with respect to the input sample density) onto a rectangular grid.
    \item \textbf{Rect + Bic ($\mathbf{4\times}$).} For rectangularly sampled data, we use bicubic interpolation (Bic) to upscale by a factor of 4.
    \item \textbf{Hex + NI ($\mathbf{4\times}$) + Wiener.} To the Hex + NI (x4) system, we add Wiener filtering\cite{Gonzalez2006} for restoration.
    \item \textbf{Rect + Bic ($\mathbf{4\times}$) + Wiener.} To the Rect + Bic ($4\times$) system, we add Wiener filtering for restoration.
    \item \textbf{Hex + NI ($\mathbf{2\times}$) + RCAN ($2 \times$).} As proposed in Sec.~\ref{sec:approach}, for hexagonally sampled data, we use NI to upscale by a factor of 2 (with respect to sample density) onto a rectangular grid and then feed the imagery into an RCAN ($2 \times$) network. (See Fig.~\ref{fig:flowchart}(a).)
    \item \textbf{Rect + Bic ($\mathbf{2\times}$) + RCAN ($2 \times$).} To mirror Hex + NI ($2\times$) + RCAN ($2 \times$) with rectangularly sampled data, we use bic to upscale by a factor of 2 and then feed the imagery into an RCAN ($\times 2$) network. (See Fig.~\ref{fig:flowchart}(c).)
    \item \textbf{Rect + RCAN ($4 \times)$.} For rectangularly sampled data, we use an RCAN ($4 \times$) network. (See Fig.~\ref{fig:flowchart}(b).)
\end{itemize}
Nonuniform interpolation is implemented with MATLAB's scatteredInterpolant function and bicubic interpolation is implemented with MATLAB's interp2 function.
For the systems that use Wiener filtering, one Wiener filter\cite{Gonzalez2006} is used for each RGB channel using the observation model's PSF for that channel.
The noise-to-signal ratio (NSR) parameter of the Wiener filters are fit by finding an NSR parameter between $10^{-5}$ and $1$ that minimizes mean squared error (MSE) between the output of the Wiener filter and the desired image for every image in the training data (independently for each channel).
The per-channel NSR values are averaged across the training data.
The training methodology for RCAN networks and the evaluation protocol are described in the rest of this section.

\subsection{Data and preprocessing}

For data, we require a high-resolution dataset with which to simulate realistic camera imagery using the observation model described in Sec.~\ref{sec:obs_model}.
We choose the DIV2K dataset\cite{Agustsson_2017_CVPR_Workshops}, which consists of 2K resolution natural imagery.
The full DIV2K dataset consists of 800 training images, 100 validation images, and 100 testing images.
However, SR truth for the testing images has not been publicly released.
Following Ref.~\citenum{Lim2017EnhancedDR}, we report out testing performance using a portion of the DIV2K validation set.
In particular, our train split consists of DIV2K images 1-800, our validation split consists of DIV2K images 801-810, and our test split consists of DIV2K images 811-900.
This approach ensures a clean separation of data for training, validation, and testing.

During preprocessing, the scatteredInterpolant MATLAB function is used to resample the hexagonally imagery to the rectangular grid in preprocessing for NI systems and the interp2 MATLAB function is used to resample rectangularly sampled imagery for bic systems.
We replace any NaN values that arise during this step with ``0.''
For data augmentation, random left-right flip, random up-down flip, and transposing are used to generate all eight possible combinations of flipping and rotation.
Data augmentation is applied \emph{before} applying the observation model.
That is, for each image in the original dataset, we transform it into eight images.
The observation model is applied to each transformed image separately.
As a result of this, each epoch of training for experiments with data augmentation consists of a pass through \emph{all of the augmentation sets}, meaning every image in the DIV2K training images is seen 8 times per epoch.

\subsection{Training hyperparameters}

Our RCAN networks are trained with default settings from the original RCAN network: we also use 10 residual groups and 10 residual blocks.
We consider both randomly initializing our networks and initializing our networks with publicly available pre-trained RCAN ($2 \times$) and RCAN ($4 \times$) network parameters.
These pre-trained networks were originally trained to upsample images that were downsampled with bicubic interpolation.
It should be noted that it is not a priori obvious that using pre-trained weights is effective.
We chose to consider pre-training because the original RCAN networks found that initializing the networks RCAN ($3 \times$) and RCAN ($4 \times$) with the pre-trained parameters from RCAN ($2 \times$) is effective, showing that the weights from one model are broadly applicable to slightly different scales.
We consider whether a similar intuition applies when the scales are held constant but the task is slightly changed (namely, realistic degradations are applied to the imagery).
This choice is considered in the ablation study in Sec.~\ref{sec:ablation}.
The sub-pixel distance matrix input head described in Table~\ref{tab:dist_head} is randomly initialized in all cases. 

A patch size of $96 \times 96$ pixels is used for training.
The batch size is set to 16 (patches).
The Adam optimizer\cite{kingma2017adam} is used with a learning rate of $10^{-4}$.
The hexagonal model is trained for 200 epochs (1600 passes through the data).
The learning rate is decayed by a factor of $0.5$ at steps 120 and 160.
It is found that a much shorter training schedule is appropriate for the rectangular model.
By adjusting the schedule based on the validation learning curve, we arrived at training length of 23 epochs with decay at steps 8 and 15.
Presumably, this is because the rectangular model, initialized with pre-trained weights, needs only to learn to adapt to the optical degradation whereas the hexagonal model needs to also undo resampling noise from the process interpolating rectangularly gridded samples.

Network training and testing is performed with an Ubuntu 18.04 workstation with an AMD Ryzen 7 2700X Eight-Core Processor with a maximum clock speed of 4 GHz. It is equipped with two NVIDIA GeForce RTX 2080 Ti Graphics Processing Units (GPUs). Both GPUs are used during training. Training the RCAN ($2\times$) models as described takes approximately one day. Training the RCAN ($4\times$) model takes approximately four hours. Testing on our 90 image test set takes approximately one minute.

\subsection{Loss function}\label{sec:loss_function}

\begin{figure}[ht]
    \centering
    \includegraphics[clip=true, trim = .5in 3in .5in 3in, width=5.5in]{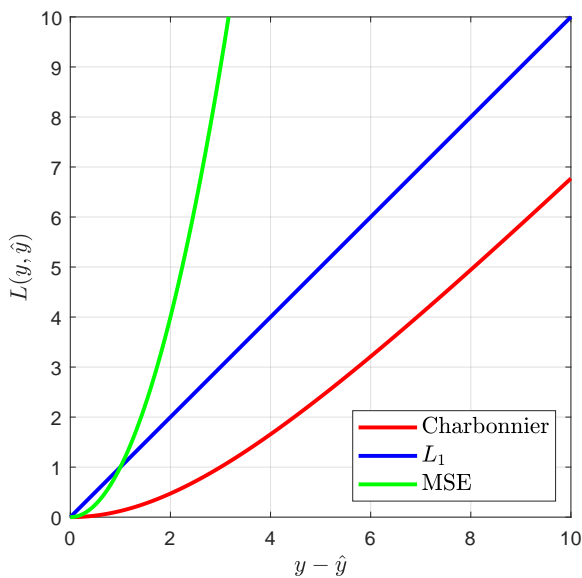}
    \caption{Plot of the Charbonnier loss as a function of the error term $y - \hat{y}$ with $\varepsilon=4$ (red), the L1 loss function (blue), and the MSE loss function (green).}
    \label{fig:charbonnier}
\end{figure}

We consider three loss functions: MSE, 
L1 loss, and the Charbonnier loss function.
For a loss function, the traditional choice is MSE, since this effectively 
seeks to maximize the signal-to-noise ratio (SNR).
The formula for MSE is given by
\begin{equation}
    L(y,\hat{y}) = \text{mean}((y - \hat{y})^2),
\end{equation}
where $y$ is the true image, $\hat{y}$ is the estimated image, the squaring is performed elementwise, and $\text{mean}$ is taken across all pixels in each batch.  
While MSE is theoretically sound, empirical studies in Ref.~\citenum{Lim2017EnhancedDR} found the L1 loss to be superior to MSE for training EDSR and the original RCAN networks were trained with an L1 loss \cite{Zhang2018ImageSU}.
The formula for L1 loss is given by
\begin{equation}
    L(y,\hat{y}) = \text{mean}(|y - \hat{y}| ).
\end{equation}
While MSE and L1 are common, we are inspired to consider a Charbonnier loss function 
used in the joint deblurring and super-resolution model ED-DSRN \cite{Zhang2018ADE}. 
This loss function has been shown to perform well in a task that is similar to the one at hand here.  The formula for the Charbonnier loss function is
\begin{equation}
    L(y,\hat{y}) = \text{mean}(\sqrt{(y - \hat{y})^2 + \varepsilon^2} - \varepsilon).
\end{equation}
A visual comparison of the Charbonnier loss with the L1 loss function and the MSE loss function is given in Fig.~\ref{fig:charbonnier}.
The advantage of the Charbonnier loss function is that the function is smooth when error is close to 0 like MSE, but it grows linearly when the error is large like L1.  We believe this
improves global convergence during training.
In our application, the intensity values of images $y$ and $\hat{y}$ are scaled to an approximate range of $0$ to $255$. For this range we have observed good results using $\varepsilon=4$, and this value is used in all of our experiments.

\subsection{Evaluation criteria}\label{sec:metrics}

We consider two evaluation metrics: peak signal-to-noise ratio (PSNR) and Structural Similarity (SSIM) index \cite{Wang2004ImageQA}.
We consider both metrics in two settings: First, we convert the image to grayscale by evaluating the Y channel of the YCbCr color space and evaluate our metrics.
Second, we use full-color RGB imagery.
In all cases, we shave 6 pixels from each edge of the image to avoid evaluating border effects.
Note that our results are not directly comparable with other prior results in the literature because of the use of the
camera-parameter based observation model employed here\cite{Elwarfalli2021FIFNETAC}.  
We believe the observation model provides more realistic data, but it is not yet widely used in the literature.
Note we report PSNR values in dB.  

\section{Experimental Results}\label{sec:exper_results}

In this section, we describe the results of the hexagonal and rectangular image processing systems on our test split (DIV2K images 811-900) as well as provide an ablation study of our architectural choices.
We provide both quantitative and qualitative analysis.
We consider the saved checkpoint with the best validation PSNR on the Y channel.

\subsection{Quantitative Results}

We calculate the PSNR and SSIM as described in Sec.~\ref{sec:metrics}.
The results are given in Table~\ref{tab:test_results}.

\begin{table}[ht]
    \caption{Results on our test split (DIV2K images 811-900). ``NI'' refers to nonuniform interpolation and ``Bic'' refers to bicubic interpolation.  The bold face entries represent the best results for the given metric.}
    \label{tab:test_results}
    \centering
    \begin{tabular}{l|cc|cc|cc|cc}
    & \multicolumn{2}{c|}{Y channel} & \multicolumn{2}{c|}{RGB} \\
      & PSNR (dB) & SSIM & PSNR (dB) & SSIM \\
\hline
        Hex + NI ($4\times$) & 25.424 & 0.70895 & 23.944 & 0.69016 \\
        Hex + NI ($4\times$) + Wiener & 26.460 & 0.75731 & 24.958 & 0.73770 \\
        Hex + NI ($2\times$) + RCAN ($2 \times$) & $\mathbf{29.204}$ & $\mathbf{0.84425}$ & $\mathbf{27.692}$ & $\mathbf{0.82957}$\\
        \hline
        Rect + Bic ($4\times$) & 25.826 & 0.72621 & 24.348 & 0.70796 \\
        Rect + Bic ($4\times$) + Wiener & 
        26.060 & 0.73410 & 24.587 & 0.71642 \\
        Rect + Bic ($2\times$) + RCAN ($2 \times$) & $29.004$ & $0.83735$ & 27.494 & 0.82254 \\
        Rect + RCAN ($4 \times$) & $29.073$ & $0.83939$ & 27.556 & 0.82442 \\
    \end{tabular}
\end{table}

On our test split, our proposed Hex + NI ($2 \times$) + RCAN ($2 \times$) shown in Fig.~\ref{fig:flowchart}(a) achieves a PSNR of 29.204 and an SSIM of 0.84425 when evaluated on the Y channel and 27.692 and 0.82957, respectively, when evaluated on full RGB.
The equivalent system using rectangularly sampled imagery, Rect + Bic ($2 \times$) + RCAN ($2 \times$), achieves a PSNR of 29.004 and an SSIM of 0.83735 on the Y channel and 27.494 and 0.82254, respectively, when evaluated on full RGB.
The purely RCAN-based Rect + RCAN ($4 \times$) system shown in Fig.~\ref{fig:flowchart}(b) achieves a PSNR of 29.073 and an SSIM of 0.83939 on the Y channel and 27.556 and 0.82442, respectively, on full RGB.
This means that the hexagonal sampled imagery outperforms the rectangular sampled imagery, whether the rectangular imagery is processed by the end-to-end deep learning Rect + RCAN ($4 \times$) system ($+.131$ PSNR and $+.00486$ SSIM on Y channel, and $+.136$ and $+.00515$, respectively, on RGB) or by the two-stage Rect + Bic ($2 \times$) + RCAN ($2 \times$) system that mirrors the hexagonal processing ($+0.200$ PSNR and $+.00690$ SSIM on Y channel, and $+.198$ and $+.00703$, respectively, on RGB).
Moreover, this shows that for rectangular imagery, using only RCAN as in the Rect + RCAN ($4 \times$) system is superior to using bicubic interpolation with RCAN as in the Rect + Bic ($2 \times$) + RCAN ($2 \times$) system ($+0.069$ PSNR and $+.00204$ SSIM on Y channel, and $+0.062$ and $+.00188$, respectively, on RGB).
To ground these differences, note that the qualitative results shown in Fig.~\ref{fig:image823} represent a difference of $0.281$ PSNR and $0.008896$ SSIM between the center and right columns, which are similar magnitudes as these averaged results.
As can be seen in Table~\ref{tab:test_results}, the trends observed in the quantitative Y channel results are very similar to the trends in the RGB results, so we focus on Y channel results for the remainder of this section.

Comparing our RCAN systems against the baseline interpolation and interpolation + Wiener systems in Table~\ref{tab:test_results}, our RCAN-based post-processing achieves higher performance.
Every RCAN-based system outperforms every baseline system by at least $+2.544$ PSNR and $.08004$ SSIM.  Comparing baseline systems across sampling type in Table~\ref{tab:test_results}, we find that $4 \times$ interpolation of rectangularly sampled imagery is better than $4 \times$ interpolation of hexagonally sampled imagery.
To be precise, Rect + Bic ($4 \times$) outperforms Hex + NI ($4 \times$) by $.402$ PSNR and $0.01726$ SSIM.
However, post-processing with a Wiener filter shows the benefits of hexagonally sampled imagery:
Hex + NI ($4 \times$) + Wiener outperforms Rect + Bic ($4 \times$) + Wiener by $.400$ PSNR and $.02321$ SSIM.

\subsection{Qualitative Results}

\begin{figure}
    \centering
    \includegraphics[width=\textwidth]{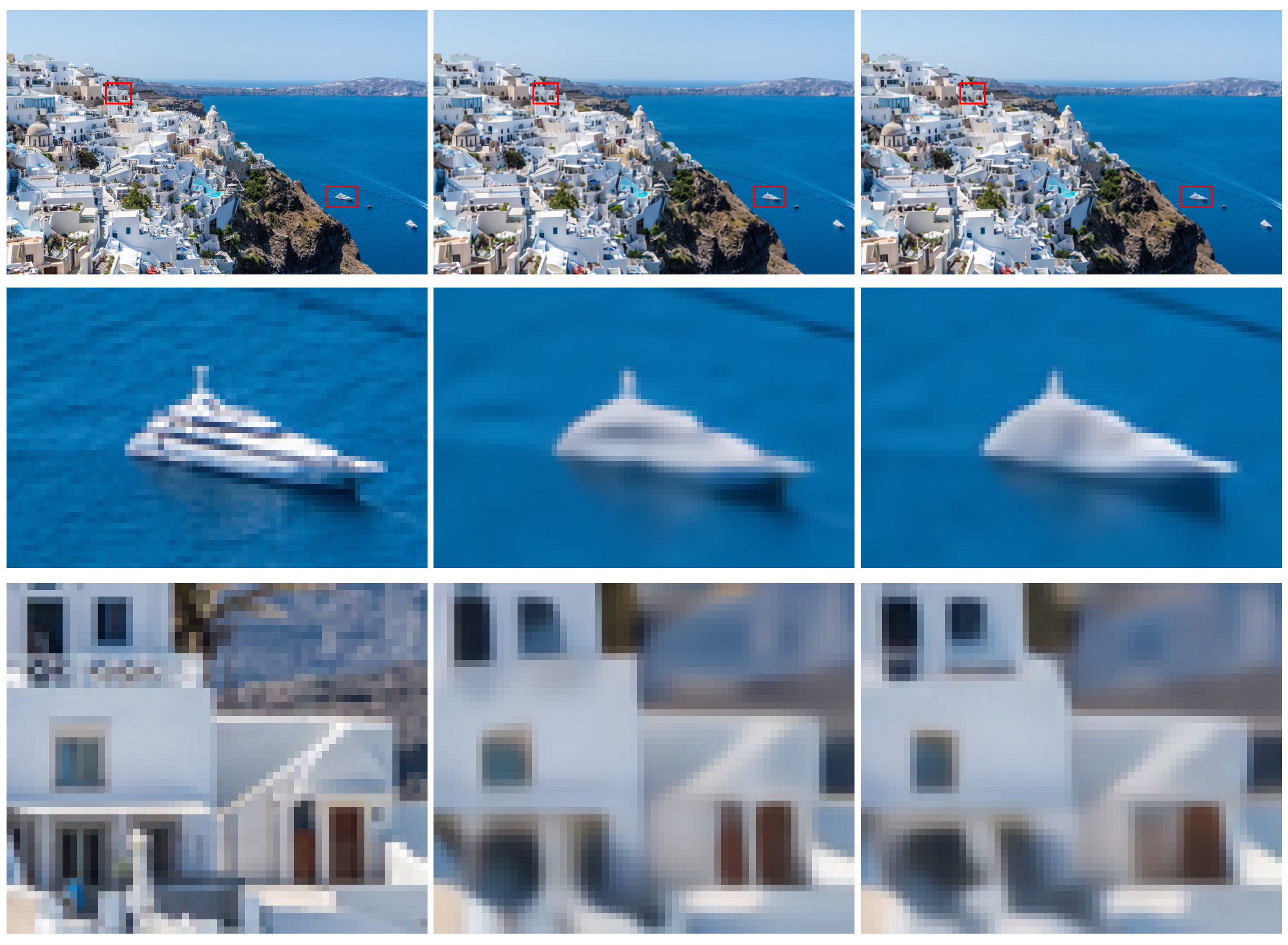}
    \caption{DIV2K Image 823. The truth is shown in the left column, the output of
    Hex + NI ($2 \times$) + RCAN ($2 \times)$ processing is shown in the middle column, and Rect + RCAN ($4 \times$) outputs are shown in the rightmost column.  On the first detail row, the hexagonal model captures some notion of a horizontal line corresponding to the decks of the ship, which is absent from the rectangular model. On the second detail row, the hexagonal model renders the brown door better than the rectangular model.}
    \label{fig:image823}
\end{figure}

\begin{figure}
    \centering
    \includegraphics[width=0.9\textwidth]{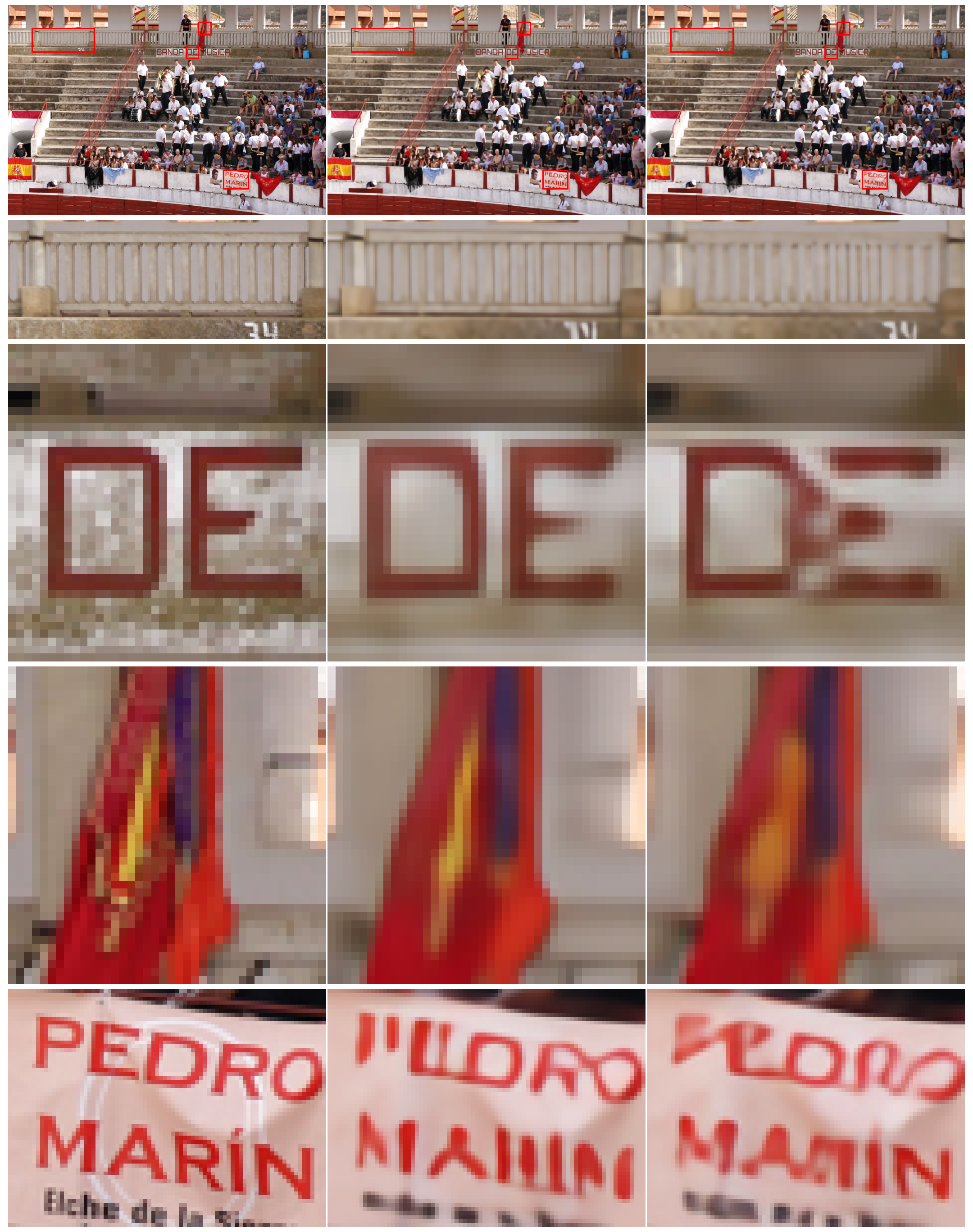}
    \caption{DIV2K Image 825. On the first detail row, the ``D'' and ``E'' are rendered more faithfully by the hexagonal model than the rectangular. On the second detail row, the yellow band in the flag is rendered more faithfully by the hexagonal system. On the final detail row, the top row of letters (particularly the ``P,'' ``D,'' ``R,'' and ``O'') are more faithfully rendered by the hexagonal system.}
    \label{fig:image825}
\end{figure}

\begin{figure}
    \centering
    \includegraphics[width=0.7\textwidth]{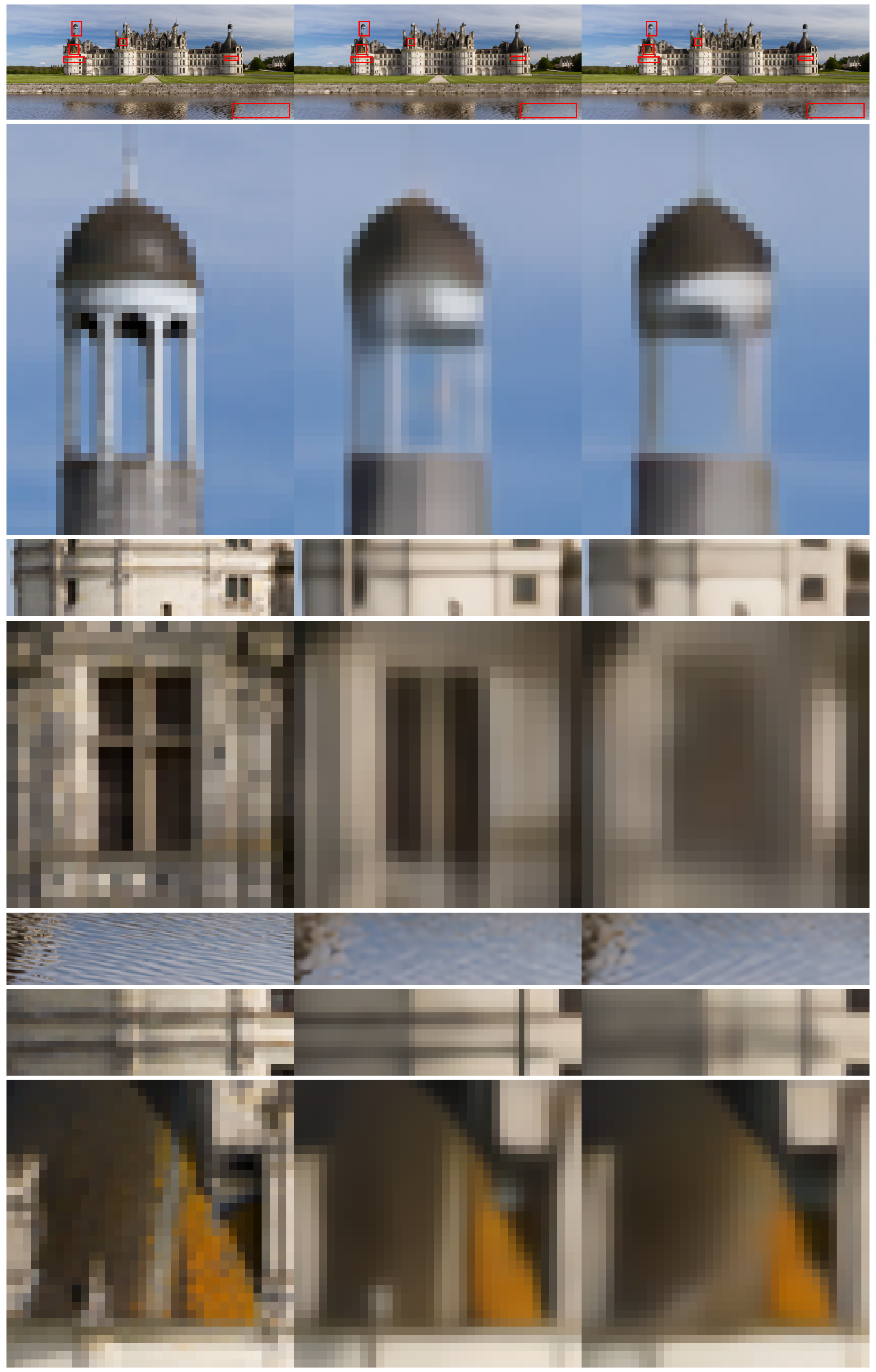}
    \caption{DIV2K Image 830. On the first detail row, the hexagonal model captures all vertical supports while the rectangular does not. On the second detail row, the rectangular model mis-colors the band between the lines on the building while the hexagonal model does not. On the third detail row, the hexagonal model captures the vertical support in the window while the rectangular model renders a blur. On the fourth detail row, the hexagonal model generates interference patterns in the water whereas the rectangular model is dominated by one direction of line. On the fifth detail row, the hexagonal rendering is clear whereas the rectangular rendering is blurry. On the sixth detail row, the hexagonal model captures more detail between the orange and the gray areas than the rectangular model.}
    \label{fig:image830}
\end{figure}

\begin{figure}
    \centering
    \includegraphics[width=\textwidth]{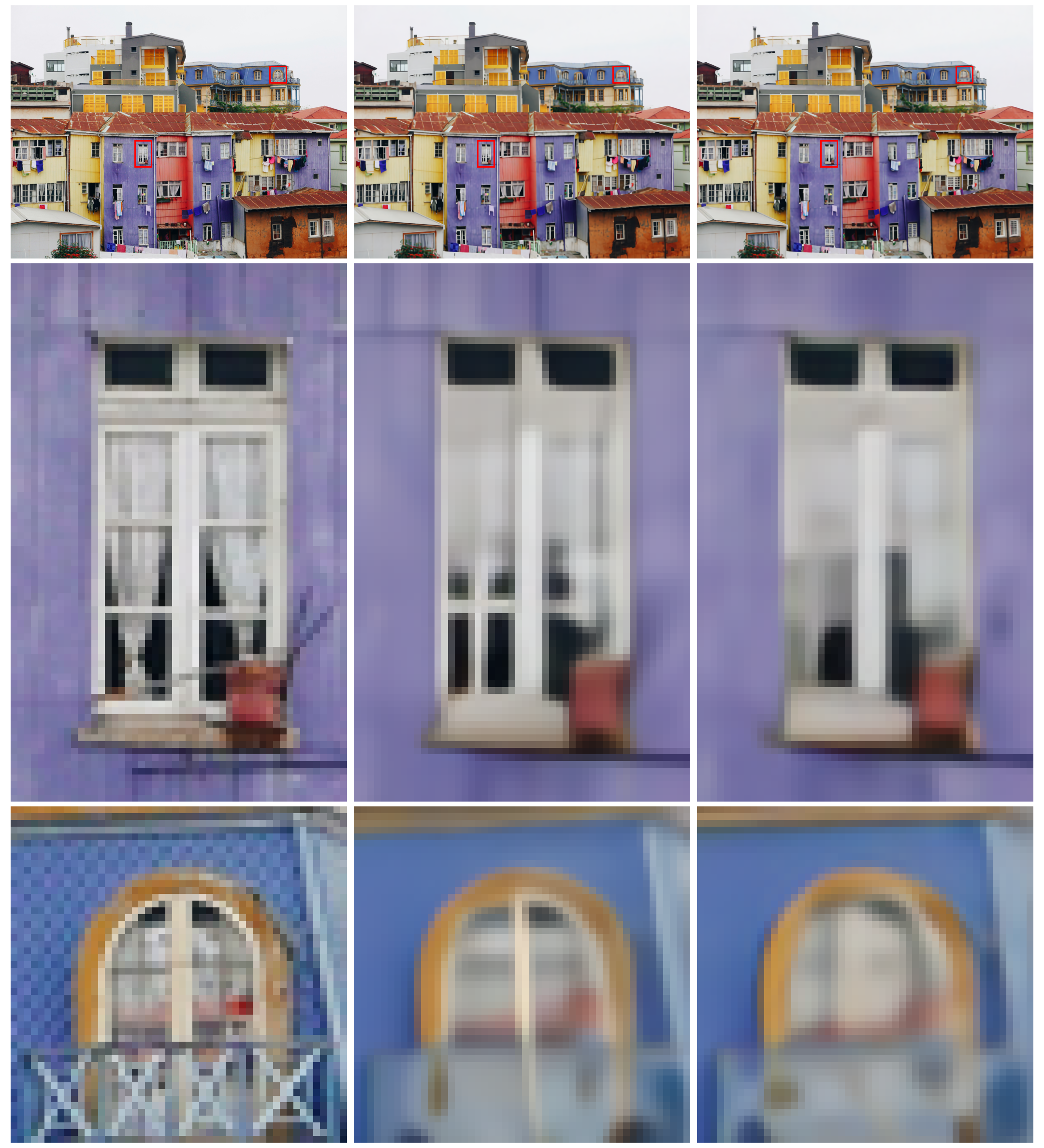}
    \caption{DIV2K Image 836. On both rows, the hexagonal model captures more detail of the vertical and horizontal supports of windows.}
    \label{fig:image836}
\end{figure}

\begin{figure}
    \centering
    \includegraphics[width=0.8\textwidth]{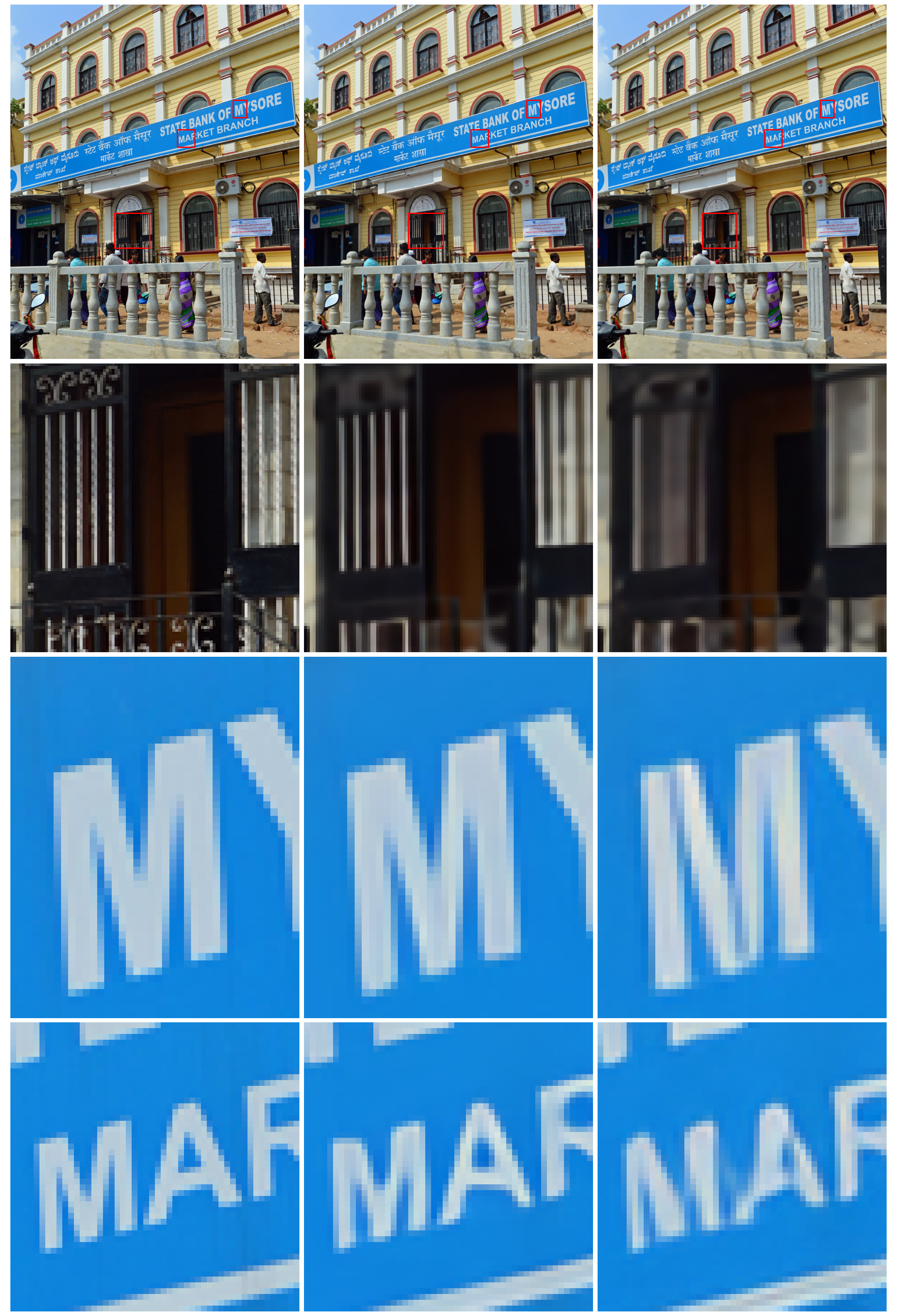}
    \caption{DIV2K Image 891. In the first detail row, the hexagonal model captures the detail of vertical supports in a window more faithfully than the rectangular model. In the second and third detail rows, the hexagonal model more faithfully renders letters in text.}
    \label{fig:image891}
\end{figure}

\begin{figure}
    \centering
    \includegraphics[width=\textwidth]{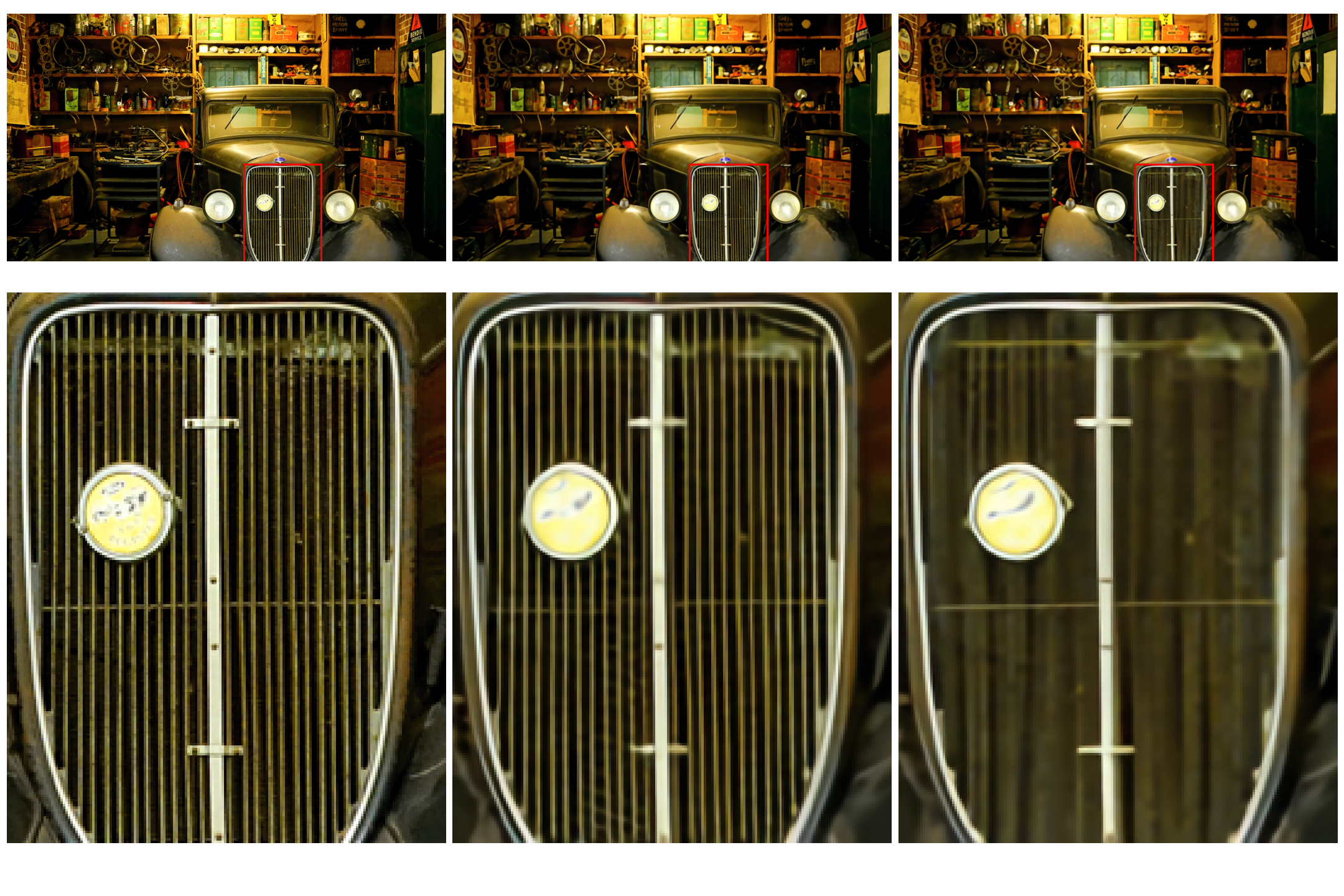}
    \caption{DIV2K Image 900. The hexagonal system faithfully renders the vertical lines in the grill of the car whereas the rectangular image is clearly aliased.}
    \label{fig:image900}
\end{figure}

For a qualitative comparison between our proposed Hex + NI ($2 \times$) + RCAN ($2 \times)$ system and the Rect + RCAN ($4 \times$) system, close-ups of images from our test split are provided in Figs.~\ref{fig:image823}-\ref{fig:image900}.
In all of these figures the truth is shown in the left column, the output of
Hex + NI ($2 \times$) + RCAN ($2 \times)$ processing is shown in the
middle column, and Rect + RCAN ($4 \times$) outputs are shown in the rightmost column.
The full images are shown in the top row and regions of interest (ROIs) are highlighted 
in subsequent detail rows.  Themes throughout these images include more faithful rendering of text (Figs.~\ref{fig:image825}, \ref{fig:image891}), improved rendering of support beams in small windows and doors (Figs.~\ref{fig:image823}, \ref{fig:image830}, \ref{fig:image836}), and more faithfully capturing textures with high-spatial-frequency lines (Figs.~\ref{fig:image891}, \ref{fig:image900}).

\subsection{Ablation Study}\label{sec:ablation}

In this section, we present results from our ablation study.
Since these experiments were actually carried out during the model selection phase, only Y channel scores on our validation split (DIV2K images 801-810) are reported.

\subsubsection{Hexagonal system}

For the hexagonal model's ablation, we are interested in the effects of the loss function, the sub-pixel distance matrix, the augmentation scheme, and the choice of pre-trained weights (scratch vs RCAN ($2 \times$) vs RCAN ($4 \times$)).
The results of these experiments are described in Table~\ref{tab:hex_ablation}.

Regarding choice of the loss functions described in Sec.~\ref{sec:loss_function}, we find that the Charbonnier loss function is the most effective.
The L1 loss function imposes a penalty of 0.004 in PSNR for a score of 29.578.
The MSE loss function is found to be much worse, imposing a penalty of 0.031 in PSNR for a score of 29.551.
For the remainder of the experiments, we consider the Charbonnier loss function.

Going down the rows of Table~\ref{tab:hex_ablation}, in the final block we successively remove features from the architecture.
First, we remove the distance head described in Sec.~\ref{sec:sub_pixel_distance}.
A penalty of 0.003 in PSNR is observed for a score of 29.579.
Next, we do not use data augmentation and adjust the schedule somewhat to account for the fact that each epoch now contains fewer images (see Sec.~\ref{sec:training_details}).
A further penalty of 0.132 in PSNR is observed for a score of 29.447.

Next, we consider the effects of using pre-trained weights as discussed in Sec.~\ref{sec:training_details}.
First, we consider using the pre-trained RCAN ($4 \times$) weights instead of the pre-trained RCAN ($2 \times$) weights.
This tests the effect of using \emph{some} pre-trained weights, even if the weights are sub-optimal.
Note the pixel shuffle layer is randomly initialized for this experiment since the $4 \times$ pixel shuffle layer would return the wrong image size.
A penalty of $0.015$ in PSNR is observed from the last step in the ablation for a score of $29.432$.
This stands in contrast to no pre-training at all, where a further penalty of $0.221$ in PSNR is observed for a score of $29.211$.
So, using sub-optimal pretrained weights is found to be better than training from scratch.
This is consistent with the result from Ref.~\citenum{Zhang2018ImageSU}, where using one scale of RCAN weights was found to be effective for initializing a training for another scale.

\begin{table}
    \centering
    \caption{Hexagonal system ablation study.}
    \label{tab:hex_ablation}
    \begin{tabular}{l|c|cccc|c}
        \hline\hline
        Name         & Schedule          & PT & Aug & Dist & Loss  & PSNR (dB) \\
        \hline\hline
        Hex          & \{120, 160, 200\} & x2 & Yes & Yes      & Charb & 29.582 \\
        \hline
        Hex L1       & \{120, 160, 200\} & x2 & Yes & Yes      & L1    & 29.578 \\
        Hex MSE      & \{120, 160, 200\} & x2 & Yes & Yes      & MSE   & 29.551 \\
        \hline
        -Dist        & \{120, 160, 200\} & x2 & Yes & No        & Charb & 29.579 \\
        -Aug         & \{300, 400, 501\} & x2 & No  & No        & Charb & 29.447 \\
        -x2+x4       & \{300, 400, 501\} & x4 & No  & No        & Charb & 29.432 \\
        -PT          & \{300, 400, 501\} & No & No  & No        & Charb & 29.211 \\
        \hline\hline
    \end{tabular}
\end{table}

To recap the results of the hexagonal ablation study, we find performance drops the most when we remove augmentation ($-0.132$ in PSNR) and when we train from scratch ($-0.221$ in PSNR).
Consequently, we believe that using pre-trained weights (despite the weights not representing how to denoise the image) and using data augmentation are especially important to our success.
We note, however, that our adaptation of the training schedule for data augmentation is imperfect, which makes it more difficult to directly compare experiments before and after removing data augmentation.
A better schedule for non-augmented training would be $\{960, 1280, 1600\}$ to match the number of examples seen in total during training with augmentation.
However, since less data is seen overall, such a long schedule is not necessarily required in practice and we do note that the learning curve still appears to be mainly converged.

\subsubsection{Rectangular system}

Next, we perform an ablation study for our rectangular system.
Note that we only perform ablation for Rect + RCAN ($4 \times$) and not Rect + Bic ($2 \times$) + RCAN ($2\times$) because it provided the best performance. 
Here, there are fewer design decisions to be made.
In particular, there is no use for a sub-pixel distance matrix since every sample is perfectly aligned with the sample grid (i.e., there is no interpolation noise).
On the other hand, we consider the effects of shortening the training schedule (see Sec.~\ref{sec:training_details}).
The ablation study is summarized in Table~\ref{tab:rect_ablation}.

\begin{table}
    \centering
    \caption{Rectangular model ablation study.}
    \label{tab:rect_ablation}
    \begin{tabular}{l|c|ccc||c}
        \hline\hline
        Name         & Schedule          & PT & Aug & Loss  & PSNR (dB) \\
        \hline\hline
        Rect         & \{8, 15, 23\}     & x4 & Yes & Charb & 29.556 \\
        \hline
        +Long        & \{120, 160, 200\} & x4 & Yes & Charb & 29.543 \\
        \hline
        -x4+x2       & \{8, 15, 23 \}    & x2 & Yes & Charb & 29.446 \\
        -PT          & \{120, 160, 200\} & No & Yes & Charb & 29.306 \\
        -Aug         & \{300, 400, 501\} & No & No  & Charb & 29.010 \\
        \hline\hline
    \end{tabular}
\end{table}

Going down the rows of Table~\ref{tab:rect_ablation}, we first consider the effect of training schedule.
When we train for the same length as the hexagonal model, a penalty of 0.013 in PSNR is observed for a score of 29.543.
Next, we consider the effects of pre-training in the same manner as with the hexagonal model.
When we swap to sub-optimal weights, a penalty of $0.097$ in PSNR is incurred for a score of 29.446.
When we train from scratch, a further penalty of $0.140$ in PSNR is incurred for a score of 29.306.
Finally, when we remove augmentation, we incur another penalty of 0.296 in PSNR for a score of 29.010.
Again, we find that using data augmentation and using some form of pre-training are especially significant to our performance.

\section{Discussion}\label{sec:ccl}

We demonstrate an approach to resampling hexagonal imagery to a rectangular grid that retains the benefits of hexagonally sampled imagery.
This approach uses
non-uniform interpolation to resample the 
hexagonal data onto an upsampled rectangular grid.  We then apply RCAN, a state-of-the-art deep learning-based super-resolution system, for restoration and further SR processing.
To aid the RCAN processing, we also 
code the distance of each interpolated sample to the nearest original hexagonal sample
and provide this as auxiliary input features to RCAN.
To demonstrate the approach, we use a camera degradation model that simulates diffraction and detector integration of a hexagonal camera given a high-resolution rectangularly sampled "truth" image.

On the DIV2K SR dataset, we find that the hexagonal system outperforms applying RCAN directly on rectangularly sampled imagery with the same sample density, no interpolation noise, and modeled rectangular camera degradations in terms of PSNR and SSIM.
Images from the holdout set are provided that show clearly visible differences between the outputs generated by the hexagonal system and by the rectangular system.
We also find that the hexagonal system outperforms an identical two-stage system that ingests and is trained on realistic rectangular imagery.

Theoretically, hexagonal sampling encodes more frequency information.
We posit that this theoretical benefit of hexagonal sampling is significant for our success in achieving superior SR performance.
We would expect to see some amount of benefit to hexagonal sampling  for a wide range of sensor parameters, provided that the corresponding rectangular camera system is undersampled.
However, we live in a predominantly rectangularly sampled world: computer monitors, cell phone screens, and printers all usually use rectangularly sampled images.
As a result, the typical tools for processing imagery assume rectangular sampling grids.
We believe that our system provides a bridge between hexagonal samples and rectangular processing that reduces the loss incurred by converting to the rectangular sampling scheme.

Future work should consider directly learning to jointly resample, restore, and increase the resolution of the imagery within one end-to-end deep learning system.
Since we report that the rectangular system based on RCAN ($4 \times$) outperforms the two-stage rectangular system significantly, we posit that an end-to-end system for hexagonal imagery will be similarly more performant.
Developing such a system is a challenging problem because there is no clear-cut method for resampling the hexagonal imagery to a rectangular grid with convolutions or a PixelShuffle layer.
Theoretically, however, such a method is advantageous because it avoids introducing interpolation noise in the resampling step.
Hopefully the availability of libraries like HexagDLy \cite{hexagdly_paper} will inspire future researchers to pursue this problem.

\subsection*{Disclosures}
That authors declare that there are no relevant financial interests in the manuscript and no other potential conflicts of interest to disclose.

\subsection* {Code, Data, and Materials Availability} 
The truth dataset used in our study is publicly available and described in Ref.~\citenum{Agustsson_2017_CVPR_Workshops}.
The author's code and degraded image are not publicly available at this time, but are available from the authors upon reasonable request.


\bibliography{article}   
\bibliographystyle{spiejour}   


\vspace{2ex}\noindent\textbf{Dylan Flaute} graduated {\em magna cum laude} with a Bachelor's degree in
electrical engineering and a Bachelor's degree in mathematics from the University of Dayton in December 2020. 
There, he was awarded The Thomas R. Armstrong, '38 Award of Excellence for Outstanding Electrical Engineering Achievement from the School of Engineering and The Senior Award for Academic Excellence in Mathematics from the Department of Mathematics.
He is currently a Junior Autonomy Engineer at Jacobs, where he works with the Air Force Research Laboratory's ACT3 team.
His research interests include digital signal and image processing and machine learning. 

\vspace{2ex}\noindent\textbf{Russell C. Hardie} is a Full Professor in Department of Electrical and Computer Engineering at the University of Dayton, and holds a joint appointment in the Department of Electro-Optics and Photonics.  Dr. Hardie received the University of Dayton's top university-wide teaching award in 2006 (the Alumni Award in Teaching).  He also received the Rudolf Kingslake Medal and Prize from SPIE in 1998 for work on super-resolution imaging. His research interests include a wide range of topics in signal and image processing, image restoration, medical image analysis, and machine learning.

\vspace{2ex}\noindent\textbf{Hamed Elwarfalli} is a Ph.D. student in the Department of Electrical and Computer Engineering at the University of Dayton, where he
received his Master of Science degree in Electrical Engineering in 2017.
He received a  Bachelor's degree in Electrical Engineering from the College of Electronics Technology, Libya, in 2010.  His research interests include image processing 
and machine learning.

\listoffigures
\listoftables

\end{document}